Effects of impact and target parameters on the results of a kinetic impactor: predictions for the Double Asteroid Redirection Test (DART) mission

Short Title: Impact Modeling Predictions for DART

Angela M. Stickle[1,2], Mallory E. DeCoster[1], Christoph Burger[3], Wendy K. Caldwell[4], Dawn Graninger[1], Kathryn M. Kumamoto[5], Robert Luther[6], Jens Ormö[7], Sabina Raducan[8], Emma Rainey[1], Christoph M. Schäfer[3], James D. Walker[9], Yun Zhang[10], Patrick Michel[11], J. Michael Owen[5], Olivier Barnouin[1], Andy F. Cheng[1], Sidney Cochron[10], Gareth S. Collins[11], Thomas M. Davison[12], Elisabetta Dotto[12], Fabio Ferrari[8], M. Isabel Herreros[7], Stavro L. Ivanovski[13], Martin Jutzi[8], Alice Lucchetti[14], Elena Martellato[15], Maurizio Pajola[14], Cathy S. Plesko[4], Megan Bruck Syal[5], Stephen R. Schwartz[16], Jessica M. Sunshine[17], Kai Wünnemann[6,18]

[1] Johns Hopkins University Applied Physics Laboratory
[2] Corresponding author: angela.stickle@jhuapl.edu
[3] Institut für Astronomie und Astrophysik (IAAT), Eberhard Karls Universität Tübingen, Germany
[4] Los Alamos National Laboratory
[5] Lawrence Livermore National Laboratory
[6] Museum für Naturkunde, Leibniz-Institut für Evolutions- und Biodiversitätsforschung, Berlin, Germany
[7] Centro de Astrobiología CSIC-INTA, Instituto Nacional de Técnica Aeroespacial, 28850 Torrejón de Ardoz, Spain
[8] Space Research and Planetary Sciences, Physikalisches Institut, University of Bern, Bern, Switzerland
[9] Southwest Research Institute, San Antonio TX
[10] Universite Côte d'Azur, Observatoire de la Côte d'Azur, Centre National de la Recherche Scientifique, Laboratoire Lagrange, Nice, France
[11] Department of Earth Science & Engineering, Imperial College, London, UK
[12] INAF-Osservatorio Astronomico di Roma, Roma, Italy
[13] INAF Osservatorio Astronomico di Trieste, Trieste, Italy
[14] INAF Osservatorio Astronomico di Padova, Padova, Italy
[15] Dipartimento di Scienze e Tecnologie, Università degli Studi di Napoli "Parthenope", Naples, Italy
[16] University of Arizona
[17] University of Maryland, College Park, MD
[18] Freie Universität Berlin, Germany

Effects of impact and target parameters on the results of a kinetic impactor: predictions for the Double Asteroid Redirection Test (DART) mission

ABSTRACT

The Double Asteroid Redirection Test (DART) spacecraft will impact into the asteroid Dimorphos on September 26, 2022 as a test of the kinetic impactor technique for planetary defense. The efficiency of the deflection following a kinetic impactor can be represented using the momentum enhancement factor, $\beta$, which is dependent on factors such as impact geometry and the specific target material properties. Currently, very little is known about Dimorphos and its material properties that introduces uncertainty in the results of the deflection efficiency observables, including crater formation, ejecta distribution, and $\beta$. The DART Impact Modeling Working Group (IWG) is responsible for using impact simulations to better understand the results of the DART impact. Pre-impact simulation studies also provide considerable insight into how different properties and impact scenarios affect momentum enhancement following a kinetic impact. This insight provides a basis for predicting the effects of the DART impact and the first understanding of how to interpret results following the encounter. Following the DART impact, the knowledge gained from these studies will inform the initial simulations that will recreate the impact conditions, including providing estimates for potential material properties of Dimorphos and $\beta$ resulting from DART's impact. This paper summarizes, at a high level, what has been learned from the IWG simulations and experiments in preparation for the DART impact. While unknown, estimates for reasonable potential material properties of Dimorphos provide predictions for $\beta$ of 1-5, depending on end-member cases in the strength regime.

1.1 The Double Asteroid Redirection Test (DART) Mission

Planetary defense is the term used to encompass all capabilities needed to identify and respond to a potential asteroid or comet impact with Earth. These activities include discovery, characterization, cataloging, and tracking of potentially hazardous asteroids (PHAs)[1]. In addition to these important observation-based activities, planetary defense activities are also focused on planning and implementing measures to deflect or disrupt an object if it were on an Earth impact trajectory.

Approaches to asteroid deflection typically fall into four general categories (Board et al. 2010): civil defense, slow-push/pull techniques, kinetic impact, and nuclear detonation. Depending on the nature and size of the incoming object as well as the warning time before impact, one of these different approaches would become preferred; they are roughly listed in order of increasing impactor size and decreasing warning time. One of the most potent deflection techniques is a kinetic impactor, which is useful when the incoming object has diameters as large as several hundred meters, or up to ~ 1 km given decades of warning time (Wackler et al. 2018). Kinetic impactors are both conceptually simple and the most technologically mature deflection technique (NRC, 2010). Kinetic impactors rely on momentum transfer to deflect a threatening object: a mass is intentionally impacted into the threatening object, and the added momentum changes the orbit of the threatening object to avoid Earth. Given warning times in the decadal-scale, velocity changes on the order of mm/sec are enough to successfully perturb the orbit from a collision course with Earth, and thereby deflect an asteroid. The Double Asteroid Redirection Test (DART) is NASA's first planetary defense mission and is the first direct test of the kinetic impactor technique for asteroid deflection (Cheng et al. 2018; Rivkin et al. 2021).

The DART mission launched November 24, 2021 (06:21:02 UTC) from Vandenberg Space Force Base in California, USA. The DART spacecraft will impact the moonlet of the (65803) Didymos binary asteroid system, named Dimorphos, on September 26, 2022. DART will impact Dimorphos at ~6 km/s, changing its orbital period around Didymos by more than 10 minutes (Rivkin et al. 2021). This period change will alter the lightcurve of the Didymos system, which will be measured following impact using ground-based telescopes.

---

[1] A PHA is a near-Earth object i) whose orbit reaches a minimum intersection distance with the Earth's one of < 0.005 AU, ii) whose absolute magnitude is of 22 or brighter and that iii) is large enough to cause regional damage in the event of impact.

The DART spacecraft will also carry a 6U CubeSat (CubeSats are built to standard dimensions (Units or "U") of 10 cm x 10 cm x 10 cm) named the Light Italian CubeSat for Imaging of Asteroids (LICIACube), which is managed by the Italian Space Agency (ASI) (Dotto et al. 2021). LICIACube will be released by DART ten days before impact to provide *in situ* observations both of the DART impact and the crater ejecta plume evolution, significantly contributing to DART's planetary defense investigation. In late 2026, about four years after DART's kinetic impact, Hera, a European Space Agency (ESA) mission, will rendezvous with Didymos. Hera consists of an orbiter and two CubeSats named Juventas and Milani, which will allow for full characterization of the composition, surface and interior structures, and dynamical states of the Didymos and Dimorphos system, as well as assessment of the DART impact (Michel et al. 2018). Specifically, Hera will measure in detail the DART impact crater size and morphology and/or reshaping of the target in addition to the actual momentum transferred to Dimorphos by measuring its mass (Michel et al. 2022, in review). Data obtained by DART and LICIACube will be combined with those obtained by the ESA Hera mission, in the framework of the Asteroid Impact & Deflection Assessment (AIDA) collaboration, to offer a fully documented impact deflection experiment.

Although LICIACube will observe the DART impact first-hand later this year and Hera will ultimately characterize the system in the future, very little is currently known about Dimorphos, the specific target of the DART mission.

## 1.2    The Momentum Enhancement Factor from a Kinetic Impactor

The efficiency of deflection following a kinetic impactor can be represented using the momentum enhancement factor, β, which is a scale factor defined as the ratio of the target momentum post deflection to the incoming momentum of the impactor (in this case, the DART spacecraft). Use of the parameter β allows comparison of asteroid response and deflection efficiency across a range of particular kinetic impactor scenarios. In an inelastic collision, β would be exactly 1. However, β can be greater than 1 due to the large amount of excavated ejecta material that can result in an impulse to the target exceeding the momentum delivered by the projectile. It is often written as 1 plus the ratio of the magnitude of the ejecta momentum to the impactor momentum.

Because β is a simple scalar ratio, care must be taken when defining β for a kinetic impact of arbitrary geometry in which the vector nature of momentum transfer affects the outcome. For example, after the DART impact, the change in Dimorphos's velocity along its orbital direction will be inferred from the measured period change. This momentum change will not be aligned with the incoming DART spacecraft momentum vector, which at impact is expected to be approximately 10° off of Dimorphos's orbital velocity direction. Moreover, the ejecta direction will depend on DART's angle of impact relative to the local surface at the impact site and will most likely not be aligned with either the incoming spacecraft momentum or Dimorphos's orbital velocity (see Rivkin et al. 2021, Figure 9). The lack of alignment in the relevant velocity vectors creates some ambiguity in how β should be defined.

For analysis purposes, the DART project defines β as a scalar representing the relative momentum components along the target's surface normal at the local impact site:

$$\beta = \frac{M\Delta\vec{v}\cdot\hat{n}}{m_{sc}\vec{V}_\infty\cdot\hat{n}} = 1 - \frac{m_e\vec{V}_e\cdot\hat{n}}{m_{sc}\vec{V}_\infty\cdot\hat{n}}, \quad (1)$$

In which the spacecraft, with mass $m_{sc}$ and relative velocity $\vec{V}_\infty$ at infinity, impacts a target of mass $M$ at a point at which the outward-pointing surface normal unit vector is $\hat{n}$. The excavated material that no longer remains attached to the crater lip (ejecta) and that escapes the target, having mass $m_e$ and momentum $\vec{p}_e$ at infinity, emerges as a mass-weighted mean velocity $\vec{V}_e = \vec{p}_e/m_e$. This definition of β is consistent with previous studies (e.g., Housen and Holsapple 2012, Stickle et al. 2015, Bruck Syal et al. 2016, Feldhacker et al. 2017, Rainey et al. 2020, Stickle et al. 2020, Luther et al. in preparation), though most studies of β and its dependencies have been conducted for idealized impacts in which the impact velocity vector is perpendicular to a smooth surface, and the deflection is measured along that impact velocity. Therefore, the definition of β in Equation (1) would reduce to the simple scalar definition used in the idealized studies. In general, as long as the impact is not too far off the surface normal, and the asteroid surface is not too rough at the spacecraft scale, Equation (1) will give a similar result to what would be achieved for the idealized impact and can therefore provide utility for understanding the efficiency of the momentum transfer and extending the DART result to other kinetic impact scenarios. However, for oblique impacts, this definition of β provides much less utility for predicting the momentum transfer and understanding the asteroid material response (though other potential definitions of β would also have shortcomings in the oblique impact scenario).

An expression for β in terms of the measured post-impact deflection can also be written as (see Rivkin et al. (2021) for derivation):

$$\beta = \frac{\frac{M}{m_{sc}}\Delta V_T - \vec{V}_{\infty_{\perp n}} \cdot \hat{e}_T + V_{\infty_n} \vec{\epsilon} \cdot \hat{e}_T}{V_{\infty_n}(\hat{n}+\vec{\epsilon}) \cdot \hat{e}_T}. \quad (2)$$

In this expression, β is defined as in Equation (1), in terms of the velocity components along the surface normal vector. The unit vector of the orbital velocity, $\hat{e}_T$, represents the orbital velocity direction of Dimorphos at impact, and $\Delta V_T = \overrightarrow{\Delta V} \cdot \hat{e}_T$ is the component of Dimorphos's change in velocity along this direction. $V_{\infty_n} = \vec{V}_\infty \cdot \hat{n}$ is the component of the spacecraft velocity vector $\vec{V}_\infty$ relative to the surface normal at the impact site. $\vec{V}_{\infty_{\perp n}}$ is the vector component of the spacecraft velocity perpendicular to $\hat{n}$. The small vector $\vec{\epsilon}$ is an offset vector between the surface normal and the ejecta direction. $\vec{\epsilon}$ is perpendicular to $\hat{n}$ and has a magnitude equal to $\tan(\alpha)$, where $\alpha$ is the angle between the ejecta momentum and the surface normal. The expression for β in Equation (2) is an exact result for a general impact geometry and captures the fact that ejecta components perpendicular to the surface normal may contribute to momentum transfer along the orbital direction. All components on the right-hand side of Equation (2) will either be known at impact or measured post-impact, with the exception of the ejecta offset vector $\vec{\epsilon}$, which must be determined through numerical impact modeling (see, for example Section 3.12).

### 1.3   The DART Impact Modeling Working Group

The DART investigation team is organized into five working groups to ensure completion of the necessary measurements to meet DART's Level 1 Requirements (a detailed description of these requirements and how the team will meet them can be found in Rivkin et al. 2021). The four L1 Requirements are listed below in their official forms, with the fourth requirement having two parts:

- DART-1: DART shall intercept the secondary member of the binary asteroid (65803) Didymos as a kinetic impactor spacecraft during its September to October, 2022 close approach to Earth.
- DART-2: The DART impact on the secondary member of the Didymos system shall cause at least a 73-second change in the binary orbital period.
- DART-3: The DART project shall characterize the binary orbit with sufficient accuracy by obtaining ground-based observations of the Didymos system before and after spacecraft impact to measure the change in the binary orbital period to within 7.3 seconds (1-σ confidence).
- DART-4A: The DART project shall use the velocity change imparted to the target to obtain a measure of the momentum transfer enhancement parameter referred to as "Beta" ($\beta$) using the best available estimate of the mass of Didymos B.

- DART-4B: The DART project shall obtain data, in collaboration with ground-based observations and data from another spacecraft (if available), to constrain the location and surface characteristics of the spacecraft impact site and to allow the estimation of the dynamical changes in the Didymos system resulting from the DART impact and the coupling between the body rotation and the orbit.

Note that changes to the 'binary orbital period' mentioned here refers to changes in the orbit of the secondary around the primary, not the orbit of the binary system around the Sun. Also note that the threshold DART mission fulfills L1 Requirements 1 through 4A, and the addition of Requirement 4B constitutes the baseline DART mission.

The DART investigations cover observations, simulations, and analysis. Numerical simulations are necessary tools for providing reasonable estimates of expected results and for interpreting post-impact data once DART impacts Dimorphos. Additionally, these numerical simulations can be used to infer the physical properties of Dimorphos and further help us understand impact processes on asteroids. The AIDA/DART Impact Modeling Working Group (IWG) performs simulations of the impact and immediate aftermath using high-fidelity shock physics codes. Further information on shock physics numerical modeling codes and specific codes used by the DART IWG are provided in Section 2.

The IWG has several high-level goals in preparation for the DART impact to ensure DART meets its Level 1 requirement to obtain a measure of the momentum transfer enhancement parameter, $\beta$ (Rivkin et al. 2021). These goals are 1) to better understand the magnitude of the deflection by determining the sensitivity of impact models to impact conditions, 2) to determine the momentum transfer efficiency, $\beta$, and its sensitivity to target properties, and 3) to predict the ejecta mass and crater size following the DART impact. All three goals require numerous impact simulations and will be accomplished using a variety of numerical approaches. This paper summarizes, at a high level, what has been learned from the IWG simulations and experiments in preparation for the DART impact.

DART joins Deep Impact (DI) (A'Hearn et al. 2005) and the Lunar Crater Observation and Sensing Satellite (LCROSS) (Schultz et al. 2010, Korycansky et al. 2010) as one of three planetary-scale impact experiments. The DI impactor spacecraft crashed into the Comet 9P Tempel 1 in July 2005 at an angle of about 30° from the horizontal (A'Hearn et al. 2005, Schultz et al. 2007). While comet 9P Tempel 1 is significantly larger than Dimorphos, and the closing velocity of DI is also higher than DART's closing velocity, DI provides important information for what might be learned about target properties from these large-scale impact experiments. Prior to the DI encounter, there was a large range of possible cratering scenarios defined as possible outcomes (Schultz et al. 2005), reflecting both the uncertainties in the target properties and uncertainties in cratering physics. When the DI mission occurred, computational capabilities for simulating hypervelocity impacts were less sophisticated than what is available today. For example, the ability to simulate highly oblique impacts and use complex porosity models was not as widely available. Much of the understanding of the impact came from knowledge built from laboratory experiments (e.g., A'Hearn et al. 2005, Schultz et al. 2007, Holsapple and Housen 2007, Richardson et al. 2007). DI and LCROSS missions showed that understanding many of the features and processes recorded in laboratory experiments are scalable (within reason) and that target and impactor properties are critically important.

The past 15 years have provided tremendous improvements in the ability to simulate hypervelocity impacts, and we are able to ask and answer more sophisticated questions for DART using high-fidelity numerical simulations. The DART mission represents a controlled experiment of a kinetic impactor at planetary defense scales for which one-half of the independent variables are well understood. In this case, the projectile variables are known ahead of time, while the asteroid target variables are not. The projectile (the DART spacecraft) mass will be known to a high degree because all design parameters were controlled for and highly characterized during fabrication. The impact trajectory and velocity will be well characterized following the DART impact through data provided from spacecraft telemetry and the Didymos Reconnaissance and Asteroid Camera for Opnav (DRACO) during final minutes before impact. DART's velocity will be acquired via Doppler data using convectional radio science techniques. DART's attitude and trajectory will be initially estimated from the spacecraft's star calibration cameras and the onboard inertial measurement unit (IMU) but will be updated and refined during shape modeling efforts that will construct a global digital terrain model of Dimorphos (Daly et al. this issue). The shape modeling efforts use optical navigation techniques to more precisely determine the trajectory of the spacecraft relative to Dimorphos and to determine the location of the impact point roughly within 10 cm. Where much of the uncertainty exists is in the impact target, Dimorphos.

The Dimorphos target structure and material properties are less constrained, and that which introduces uncertainty in the results of the deflection efficiency observables, including crater formation, ejecta distribution, and the momentum enhancement factor ($\beta$). Radar observations (Naidu et al. 2016, 2020) and light curves (Scheirich and Pravec, in preparation) provide a pre-impact size estimate for Dimorphos, and post-impact shape models (Daly et al. in preparation) will provide a nominal volume estimate for Dimorphos. However, the mass is a key measurement for understanding how efficiently the target was deflected, will remain unconstrained because of uncertainties in composition and density. Additionally, material properties affecting crater and ejecta curtain formation, such as target strength, target bulk porosity, and ejecta mass, particle size, and particle velocity, are all uncertain because the exact composition of Dimorphos is unknown. Lastly, local topography will also play a large role, and be unconstrained until the delivery of DRACO images that show the local surface structure directly before impact. Therefore, in order to provide a better understanding of the effects of these unconstrained target material properties on the DART experiment, the impact modeling working group has performed simulations over a wide range of target parameters to characterize their influence on the deflection efficiency observables. These pre-impact simulation studies set the stage for activities preceding impact and provide considerable insight into how different properties and impact scenarios affect momentum enhancement following a kinetic impact. This insight provides a basis for predicting the effects of the DART impact and the first understanding of how to interpret results following the encounter.

When data are returned post-impact, the IWG will be able to better constrain the material properties of Dimorphos. These data include a measure of the orbital period change induced by DART, images of Dimorphos containing the local structure at the impact site from the DRACO imager, and images of the ejecta curtain from LICIAcube. These data will allow for additional high-fidelity impact simulations that will provide better insight into the exact value of the momentum enhancement factor following DART's impact.

In this work, we summarize results from impact simulations performed by the IWG to better understand the role of projectile and target properties coupled with variable impact conditions on the outcome of a kinetic impact. From these simulations, we can provide general predictions for what to expect, in regard to the Level 1 requirements, following the DART impact.

2. MODELING TECHNIQUES USED BY THE DART IMPACT MODELING WORKING GROUP

The AIDA/DART Impact Modeling Working Group includes an international community of impact experts and employs a three-pronged approach to inform the DART team about the impact: analytic models, hydrodynamics code simulations, and experimental investigations. Well-established analytical models (e.g.,

Holsapple 1993, Housen and Holsapple 2012) allow us to make quick calculations of crater size, ejecta mass, and momentum transfer across a wide parameter space of target strength properties (e.g., Stickle et al. 2015, Raducan et al. 2019) and provide a reliable means of getting first-order predictions. Additionally, a variety of complex hydrodynamics codes (hydrocodes) are used to simulate the DART impact in greater detail to evaluate the individual role of impact conditions and how target properties affect observable outcomes of a kinetic impactor (crater size, ejecta velocity and mass, and momentum enhancement factor, $\beta$). While computationally expensive, these codes permit the examination of physical processes at play at planetary scales, which is not possible in a laboratory environment, and allow for the tracking of specific material properties throughout the impact process (e.g., Pierazzo and Collins, 2004). Finally, experimental campaigns provide a wealth of information regarding how the deflection resulting from kinetic impactors is affected by target and projectile properties and impact parameters. Experiments provide one means of validating the more complex hydrocode simulations while also forming the basis for analytical models. Though most of the efforts undertaken by the IWG consist of numerical simulations, impact experiments provide important additional and complementary information. Descriptions of results and insights gained from experimental campaigns are included in Section 3. The AIDA/DART IWG uses all three methods to determine the outcomes of the DART impact into Dimorphos.

Using a variety of numerical and modeling approaches provides some consistency in understanding physical trends resulting from a planetary scale hypervelocity impact and also allows for exploration of a large parameter space. The IWG leverages significant community expertise through close collaboration and iteration to validate the trends discovered from the full compilation of the modeling and simulations considered in this study. This section provides a brief introduction to the use of shock physics codes for hypervelocity impact and a short description of the main codes used by the IWG.

### 2.1. Shock Physics Modeling

Hydrocodes solve the conservation equations of mass, momentum, and energy in continuous media (Meyers 1994, Benson 1992). These methods also include material equations of state (EOS) to relate physical properties and constitutive models for stress responses (Collins 2002). A variety of discretization methods exist for hydrocodes: finite-difference, finite-element, finite-volume, and Smooth Particle Hydrodynamics (SPH) (Meyers 1994, Eymard et al. 2000, Collins 2002). Finite-difference methods are pointwise and thus require a structured grid (e.g., Benson 1992). Finite-element methods rely on discrete elements (curved or rectilinear) rather than points and thus do not require a structured grid (e.g., Benson 1992). Likewise, finite-volume methods also do not require a structured grid and can be used on triangular and rectilinear grids (Eymard et al. 2000). Finite-volume methods operate by computing fluxes between cells; these methods are locally conservative (Eymard et al. 2000). Finally, SPH methods rely on points, referred to as particles, for which velocity, thermal energy, and mass are known quantities (e.g., Benz et al. 1988, Jutzi et al. 2008). These particles move freely as they are not connected to one another.

Hydrocodes can use Eulerian, Lagrangian, or Arbitrary Lagrangian-Eulerian (ALE) approaches (Anderson, 1987, Benson 1992, Collins 2002, Meyers 1994). The Eulerian approach fixes a spatial grid and allows material to flow through the fixed grid (e.g., Anderson 1987, Collins 2002). Adaptive mesh refinement (AMR) is an advanced capability to some Eulerian approaches used in studies reported here, which allows for dynamic grid refinement on areas of interest (such as material interfaces and the pressure shock wave) (Crawford 1999). The Lagrangian approach allows the mesh to deform and flow with the material (e.g., Anderson 1987, Collins 2002). The ALE approach employs a Lagrangian time step and an Eulerian remap that occurs based on user-defined conditions (Benson 1992). All of these methods have historically been used to produce reliable and robust shock codes.

Additional equations representing the behavior of materials are also necessary. Typically, a constitutive model is separated into two parts: the volumetric response of the material, summarized by a material's EOS, and the response to deviatoric strains, summarized by a strength model. The accuracy of the model predictions depends on how the material models (EOS + strength) replicate the physics of material behavior and the fidelity of the models and associated parameters. The EOS can take the form of an analytic equation, such as an ideal gas law, or a tabular form, derived from experimental data (Meyers 1994, Caldwell 2019 and references therein). Constitutive models describe the evolution of stress, strain, elastic and plastic deformation, and damage for solid materials (Meyers 2010). The availability and fidelity of EOS and constitutive models vary by code. Section 2.2 briefly describes the codes and models used by the DART IWG members and summarized in this work. Details of how crater size and momentum enhancement are calculated for these different simulations are documented in Stickle et al. (2020) and other references in this section.

2.2.    Overview of Specific Shock Physics Codes

In this section, we summarize the basics of the hydrocodes used by the IWG. Each code has been validated against experimental data in a variety of regimes, and specifically benchmarked against one another and impact experiments in the strength regime (Stickle et al. 2020). For more detailed information about some recent code validation exercises relevant specifically to planetary defense, see Appendix A. For detailed descriptions of the codes, their development, and their full capabilities, references are included in each subsection. Here, we provide short summaries to orient the reader for the following sections. For each study described within this paper, specific simulation parameters and details are described in full in the referenced papers.

Because multiple codes are used by the DART team members, it is important to understand how to compare values across different simulations. Uncertainties in the codes used in this study can arise from a variety of sources, including mesh resolution, mesh relaxers (in ALE approaches), calculating material properties in mixed-material zones, and modeling choices. Because the AIDA/DART modeling team employs a variety of numerical approaches, a benchmarking campaign was designed to better understand inherent uncertainties in period change, crater size, and expected β if different codes were used (Stickle et al. 2020, Section 2.3). The uncertainties between the results from different codes are further reduced when the porosity models and applied crush curves are closely aligned between different codes (Luther et al. in preparation, Section 2.3). Stickle et al. (2020) showed that uncertainties in material properties (and how they are represented using specific constitutive models) and target structure will have larger effects than any inherent uncertainty between different codes used to simulate the DART impact. Thus, going forward, the modeling approach of the IWG will be to span a large parameter space using all of the benchmarked codes rather than focusing on the same parameter settings for multiple codes.

### 2.2.1. CTH

CTH is a multi-dimensional, multi-material, large deformation, strong shock wave physics code developed by Sandia National Laboratories (McGlaun et al. 1990; Trucano and McGlaun 1990) that is commonly used to model impacts of projectiles into asteroid-like surfaces (e.g., Crawford 1999; McGlaun et al. 1990; Quintana et al. 2015). CTH is a two-step Eulerian finite-difference code that uses a continuum representation of materials and is massively parallelizable. The code can simulate purely hydrodynamic problems (e.g., material with no strength) or use constitutive models to simulate a strength response incorporating material properties such as pressure-dependent yield strength, damage, and porosity (using the p-α model) and includes material models appropriate for geologic materials (e.g., Crawford et al., 2013; Schultz and Crawford, 2016). A wide variety of EOS options are available, including analytical and tabular options. For the cases here, the ANalytical Equation of State (ANEOS) package (Thompson et al. 1970, Thompson et al. 1974, Thompson 1990) and Simulation-Enabled Safeguards Assessment MEthodology (SESAME) database (Lyon & Johnson 1992) are used. CTH uses adaptive mesh refinement (AMR), which improves computational efficiency while allowing the user to select areas of high-resolution to be generated within the mesh that result in better tracking of the ejecta particles, shock waves, and material interfaces (Crawford 1999). The development history and description of the models and novel features of CTH are described in full detail by McGlaun et al. and Trucano and McGlaun (McGlaun et al. 1990; Trucano and McGlaun 1990)).

### 2.2.2.    iSALE-2D/-3D

The iSALE-2D shock physics code (Wünnemann et al., 2006) is a multi-material, multi-rheology extension of the SALE hydrocode (Amsden et al., 1980), specifically developed for simulating impact processes; iSALE is similar to the older SALEB hydrocode (Ivanov et al., 1997; Ivanov and Artemieva, 2002). iSALE-3D (Elbeshausen et al., 2009; Elbeshausen and Wünnemann, 2011) uses a 3D solution algorithm similar to the SALE-2D solver, as described by Hirt et al. (1974). The development history of iSALE-3D is described in Elbeshausen et al. (2009). Both codes share the same material modeling routines, including strength models suitable for impacts into geologic targets (Collins et al., 2004) and a porosity compaction model ($\varepsilon$-$\alpha$) (Wünnemann et al., 2006; Collins et al., 2011).

### 2.2.3.    FLAG

The Free LAGrange hydrocode, developed and maintained by Los Alamos National Laboratory, employs a finite-volume ALE approach (Burton 1992, Burton 1994a, Burton 1994b). FLAG is massively parallel, has been used to model a breadth of physics (Aida et al. 2013, Fung et al. 2013, Scovel & Menikoff 2009, Black et al. 2017, Cooley et al. 2014, Tonks et al. 2007), and has been verified and validated for impact cratering applications (Caldwell et al. 2018). FLAG has also been used to model extraterrestrial asteroid impacts (Caldwell 2019, Caldwell et al. 2020, Caldwell et al. 2021). FLAG has an assortment of meshing capabilities, including AMR), as well as a number of constitutive models for solid materials (Caldwell 2019, Caldwell et al. 2018, Hill 2017, Burton et al. 2018). FLAG includes analytic EOS options and the tabular SESAME database (Lyon & Johnson 1992).

### 2.2.4. Pagosa

Pagosa (Weseloh et al. 2010) is a massively parallelized finite-difference staggered mesh Eulerian continuum mechanics code developed and maintained by Los Alamos National Laboratory for the simulation of multi-dimensional problems involving shocks in multiple materials represented in detailed geometries. Pagosa has adaptive time steps, a p-α porosity model, multiple strength models, and both SESAME (i.e., tabular) and analytical EOS capabilities. It uses an up-stream weighted, monotonicity-preserving advection scheme that conserves momentum and internal energy. PAGOSA uses a fixed grid mesh throughout the entire simulation: cell size may be varied spatially so that some regions of the mesh have a higher resolution than others, but once it is defined, the mesh cannot change over the course of the calculation.

### 2.2.5. Spheral

Spheral is an adaptive smoothed particle hydrodynamics (ASPH) code maintained at Lawrence Livermore National Laboratory. Spheral implements a number of mesh-free algorithms for modeling hydrodynamics, strong shock physics, and solid material modeling with damage and fracture. The studies presented in this work use the SPH implementation (Owen et al. 1998) extended for exact energy conservation using compatible differencing of the momentum and energy equations (Owen 2014). We employ a tensor generalization of the damage models from Benz & Asphaug (1994, 1995), described in Owen (2010). Porosity is modeled using the strain-porosity model (ε-α -- see Wünnemann et al. 2006, Collins et al. 2011). The geological material strength response uses the model of Collins et al. (2004), while for the equation of state, we use ANEOS (Thompson & Lauson 1972, Melosh 2007, Thompson et al. 2019) for geologic materials and the Livermore EOS (LEOS) library (Fritsch 2016) for most other materials.

### 2.2.6.     Bern SPH

The Bern SPH code was originally developed by Benz and Asphaug (1994, 1995) to model the collisional fragmentation of rocky bodies. This code was parallelized (Nyffeler, 2004) and further extended by Jutzi et al. (2008, 2013) and Jutzi (2015) to model porous and granular materials. The most recent version of the code includes a tensile fracture model (Benz and Asphaug, 1994, 1995), a porosity model based on the p-α model (Jutzi et al. 2008, 2009), pressure-dependent strength models (Jutzi 2015), and self-gravity.

### 2.2.7.     Miluphcuda

The SPH code miluphcuda (Schäfer et al., 2016; 2020) has been mainly developed for modeling impact and collision processes and includes various rheology models, equations of state, and self-gravity. Miluphcuda can model multiple materials. For the purposes of the DART impact simulations, the code mainly uses a Lundborg (1968) yield strength parametrization and the p-α porosity model.

2.3   The Importance of Benchmarking Impact Hydrocodes

Because of the uncertainty in target properties (e.g., porosity, material strength) and structure of Dimorphos, a large number of potential parameters must be considered. Hence, the IWG uses a wide variety of numerical simulations and multiple shock physics hydrocodes, which themselves may introduce differences in the cratering and momentum transfer. Pierazzo et al. (2008) documented variability in code results arising from how flow equations are discretized and solved, which can differ between codes. Eight numerical approaches were compared for simulating impacts into strengthless targets. This initial benchmarking campaign showed that some variability in results arose from specific solution algorithms, stability parameters within the codes, and choice of resolution. Predictions of peak pressure, crater depth, and crater diameter varied between the codes within 10-20%. Several codes in the present study were part of the original benchmarking study by Pierazzo et al. (2008) FLAG, which is present in the current study but not in the original Pierazzo study, was benchmarked using the same problems as in the Pierazzo study (Caldwell et al. 2018)

Stickle et al. (2020) performed a benchmarking campaign to compare results from several of the numerical codes used by the IWG. This campaign examined the effects of: 1) impact flow field modeling, 2) brittle failure and fracture effects, 3) target porosity effects, and 4) finite-size target effects. These comparisons provided information about the difference in expected crater size and momentum enhancement predictions based solely on code design and how material strength was represented. Because asteroids have low gravity, it is likely that the DART impact will be in the strength regime. That is, the crater size will be controlled by the strength of the material rather than the gravity of Dimorphos. Adding strength models includes an additional complication when comparing results across different hydrocodes because each code may model material behavior differently.

In general, when similar strength models were used in different codes, Stickle et al. (2020) showed predictions for crater size and momentum enhancement tend to be similar (within 15—20%); this variation is similar to what was seen in Pierazzo et al. (2008) for strengthless targets. However, if the choice of strength model is different between codes, this variation can increase significantly. Choice of material strength is also significant. Indeed, the choice of strength model and the value of strength chosen cause significantly more variation in crater size and momentum enhancement prediction than variation between codes. In all cases, the variation in momentum enhancement was larger than variation in crater size, and the reliability of the momentum enhancement prediction is highly resolution dependent.

Aside from material strength, the chosen porosity model can also affect the outcome of cratering simulations and have a similar effect on result uncertainty. A benchmarking effort (Luther et al., 2021) includes one grid-based code, iSALE-2D, and two SPH codes, Bern SPH and miluphcuda and focuses on the effect of porosity. The main scenarios model the DART impact into a homogeneous, low strength (cohesion ((i.e., the shear strength of the material at zero pressure) between 1 and 100 kPa), porous (10-50%) material. The target properties are mainly derived from regolith simulant (quartz sand and JSC-1A lunar regolith simulant; Chourey et al., 2020). Luther et al. (2021) highlights the importance of matching crush curves across different porosity models (i.e., $\varepsilon$-$\alpha$ and p-$\alpha$). Their results show generally good agreement between the codes, with $\beta$ predictions in the range of +/- 5% for cohesions below 100 kPa, which further reduces the spread between previous results (Pierazzo et al. 2018, Stickle et al. 2020). An ongoing benchmarking study with the FLAG hydrocode replicates results from Stickle et al. 2020, with similar crater dimensions and momentum enhancement factors.

3. IMPORTANT EFFECTS ON EJECTA FORMATION PROCESSES AND MOMENTUM ENHANCEMENT

A variety of parameters can affect the deflection observables following a kinetic impact, including (though not necessarily limited to) target properties such as strength, internal friction, porosity, and near-surface structure as well as impact properties such as projectile mass, projectile shape, impact velocity, and impact angle. These properties are not necessarily independent of one another and that makes the coupled response of these properties difficult to determine. For example, when the porosity of a material changes, the density and strength of the material are also affected. However, when varied individually, the effects of specific parameters can be identified and understood, which provides the basis for understanding the response when a number of different material parameters are unknown, as is the case in the forthcoming DART impact. In this section, we provide a summary of the anticipated effects on crater formation, ejecta formation, and momentum enhancement from specific material properties (identified as likely contributors to the success of a kinetic impactor) and impact parameters that are not yet fully constrained for DART.

### 3.1    Current Knowledge About Dimorphos

The Didymos binary system (Naidu et al. 2016) is classified as an S-type asteroid (Dunn et al. 2013), which is the most common type of meteorite observed to fall to Earth. S-type asteroids possess evolved compositions, with siliceous mineralogy characterized by a mixture of olivine, pyroxene and Fe-Ni metal. Dimorphos is thought to be composed of similar materials (de Leon et al. 2010); however, most of the details about Dimorphos are still unknown. The roughly 165 m diameter of the moonlet has been estimated from radar data (Naidu et al. 2020); but little is known about the structure or shape of Dimorphos. Specifically, its surface characteristics (including the presence or absence of large boulders as seen at Itokawa (Michikami et al. 2008), Bennu (Walsh et al. 2019), and Ryugu (Michikami et al. 2019)), and its internal structure and porosity remain unknown. These different morphological properties can all affect the outcome of a kinetic impact.

Although little is known about the properties of Dimorphos, inferences can be made based on other asteroids that have been visited. According to the current understanding of asteroid evolution, asteroids with diameters in the range of 200 m to 10 km are expected to have gravity-dominated rubble-pile structures, and smaller asteroids could survive with monolithic or strength-dominated structures (Walsh 2018). The size of Dimorphos is close to the crossover point of these two regimes, thereby leading to difficulties in constraining its structural and material properties.

Nevertheless, given that Dimorphos is the secondary of a binary asteroid system, it should inherit its current characteristics from its formation. Therefore, making connections to the binary formation mechanisms could shed light on target properties. Different binary formation mechanisms for a spheroidal primary have been proposed and investigated in previous studies, including: 1) Simultaneous binary formation during re-accumulation from collisional fragments; 2) Secondary formed by mass shedding from the primary-progenitor and subsequent mass re-accumulation in orbits around the primary; 3) Secondary formed by regional fission of the primary-progenitor; 4) Secondary formed by fission of a large boulder from the primary-progenitor's surface (see Walsh & Jacobson 2015 and references therein).

If Dimorphos were formed through the first three scenarios, it is most likely that the asteroid would be a rubble pile with regolith properties similar to those of Didymos. Thanks to its larger size (with a diameter ~780 m), the shape and dynamical states of Didymos are better constrained by observations (Naidu et al. 2020), and the rapid spin state of Didymos can be used to constrain its structural and material properties. Previous studies have shown that, if Didymos were a rubble pile, the critical cohesive strength to maintain its structural stability at the current observed spin would be on the order of 10 Pa with a nominal bulk density of 2170 kg/m3 (Zhang et al. 2021), which corresponds to a porosity of ~20%. The van der Waals cohesive forces would readily supply the resulting amount of cohesion, if fine regolith grains were abundant in this small body (Scheers et al. 2010). Therefore, with similar regolith properties, Dimorphos would also have a cohesion of ~10 Pa.

If Dimorphos had formed through the fourth scenario, it would likely have a monolithic structure and strength-dominated target properties. The close-up images of past space missions to asteroids showed that the surfaces of these asteroids are extensively covered by large boulders, some of which are similar in size to Dimorphos (e.g., the largest boulder on Eros has a width of ~200 m, Dombard et al. 2010; the largest boulder on Ryugu, Otohime, has a size of ~160 x 120 x 70 m, Michikami et al. 2019). Previous analyses have shown, under proper conditions, a surface boulder could be ejected into an orbit close to the primary body (Tardivel et al. 2018). In general, this scenario is less likely than the first three cases given its strict restrictions on the structural properties of the primary's progenitor.

Observations of Dimorphos by the DART and Hera missions will help to constrain formation mechanisms for the moonlet. Until the missions collect and send data, however, a wide range of potential material properties for Dimorphos are feasible. The following studies used the best current understanding of Dimorphos's material strength and porosity for modeling, which are summarized in the DART project "Design Reference Asteroid" (DRA, DART project documents, see also: Rivkin et al. 2021; Richardson et al. 2022). Some values of specific interest to the impact modeling community are summarized in Table 3. While the studies described in the following sections are not strictly limited by these values, the ranges and magnitudes of these values informed the choices of relevant parameters to study.

3.2     Deflection Parameters: Crater and Ejecta

Traditionally, impact craters and their evolution occur in either the gravity- or strength-scaling regime (Holsapple and Schmidt 1987, Schmidt and Housen 1987). This nomenclature refers to the dominant force responsible for arresting crater growth: in a gravity-scaled crater, the crater size is limited by the gravity of the target body, whereas in a strength-scaling regime, the material strength is responsible for stopping crater growth. In a hypervelocity impact, a compressive shock is generated and moves through the target after impact. This shock wave compresses material, driving it downward and outward. In regions farther from the impact point, material flows outward and upward and can be compacted or ejected. This material flow results in an expanding crater with a debris cloud (ejecta) tracking outward from the crater rim. The growth of the crater will eventually slow and stop; this process may take a few minutes for a gravity-dominated crater or as little as a few seconds in the strength-dominated case. Depending on the relative importance of gravity or strength in stopping crater growth, the crater is said to have formed in the gravity regime or the strength regime.

Impact events can cause a dynamic response in material behavior due to the effects of dynamic loading, which can complicate the predicted response of a material. For example, experiments in pumice powder reveal that the resultant crater diameters are consistent with gravity-controlled growth, even though static internal friction measurements would predict strength-controlled growth (Schultz and Gault, 1985). This is because extension behind the shock wave can reduce internal friction during excavation and allows ballistic flow (Schultz et al. 2005). In other words, the common measurement of internal friction based on static conditions (e.g., holding a vertical wall) for particulates can be overly simplistic for impacts. Therefore, we approach the treatment of the effects of target material properties in Figure 1 and Table 1 qualitatively (without showing specific static values), with the understanding that material properties are dynamic according to their inelastic response to propagation of stress waves.

In order for the cratering process on Dimorphos to be gravity-dominated, the effective strength of Dimorphos would have to be less than ~4 Pa (following scaling laws, Holsapple 1993; Stickle et al. 2015). Originally, this constraint led to the assumption that the DART impact would be in the strength regime because of the small size of Dimorphos and its low surface gravity. However, the recent small carry-on impactor (SCI) impact experiment performed by Hayabusa 2 (Saiki et al. 2017, Tsuda et al. 2020) showed that crater formation might still occur in the gravity-dominated regime, even on asteroids. The DART impact will provide an additional data point, albeit at higher impact velocity on a smaller target. The bulk of the simulation work performed by the IWG, and reported on in the paper, examines cratering in the strength regime. However, it is important to understand potential effects for gravity-dominated craters for realistic cases that may result in gravity-controlled crater growth. These include: 1) cases where the strength of Dimorphos is exceptionally low, or, 2) cases where extensional conditions exist (such as the effects of dynamic internal friction due to shock-induced extension) (Schultz et al. 2005, Schultz et al. 2013). If the cratering process is gravity-controlled following DART, the effects and expected results could be quite different than in the strength-dominated regime. For example, global deformation is one possible end-member of a gravity-controlled cratering process.

Global Deformation and the Catastrophic Disruption Conditions for Dimorphos can be described by the specific energy, Q*, needed for disruption,(Gault and Wedekind 1969, Hartmann 1969, Fujiwara et al. 1977, Jutzi et al. 2010). For a DART-sized impact, the disruption threshold is at a target diameter < 12 m, so that it is not possible for DART to disrupt Dimorphos. This concept was examined in more detail using hydrocode simulations which showed that for the current energy of the DART impact, the catastrophic disruption of the target could not be achieved (Hirabayashi et al. 2022, Richardson et al. 2022). Nevertheless, a gravity-dominated process is an end-member case to consider. Raducan and Jutzi (2022) studied DART-like impacts using the Bern SPH code and found that small impacts can significantly deform weak (<10 Pa) asteroids. For cohesionless targets (e.g., zero strength), up to 20% of the target material could be displaced, causing excavation of material from the asteroid interior, global deformation, and resurfacing. For such low cohesion targets, $\beta$ can be as high as 6. In these gravity-dominated impact processes, the resulting crater might significantly change the shape of Dimorphos (e.g., Figure 4, left).

The material properties of Dimorphos will determine whether the cratering processes resulting from the DART impact occur in a gravity-scale or strength-scale regime. The observables (crater size and ejecta dynamics) resulting from the impact will allow the investigation team to infer likely material properties of Dimorphos. In order to aid these inferences, we qualitatively summarize how specific material properties and impact parameters (e.g., impact velocity) can affect crater size and ejecta properties.

Table 1 summarizes the DART team's current best understanding of how specific properties can affect crater size and ejecta formation following the DART impact using models for both particulate materials and competent solids. In general, as strength increases, the crater size decreases; this relationship is seen using analytical scaling relationships as well (Schmidt 1987, Holsapple 1987) and is consistent across simulations, even when different "strength" parameters are assumed (e.g., cohesion v. effective strength) (Stickle et al. 2015, Prieur et al. 2017). This relationship also holds true when material internal friction, the material's ability to withstand flow or shear stress, is varied: as the friction coefficient increases, crater size decreases (Schultz et al. 2007, Prieur et al. 2017) (Figure 1).

Deconvolving the effects of porosity, cohesion (strength), and coefficient of friction on ejecta properties is not straight forward. For example, the effects of internal coefficient of friction on ejecta angle are complex. Lab scale experimental studies into sand (granular and competent sand blocks) and sandstone (dry and wet) show conflicting trends (Hermalyn and Schultz 2014, Hoerth et al. 2013). Each of these parameters can affect ejection angles differently as the crater forms. For example, ejection angles can vary at early times, but then the difference can disappear at late times. Hermalyn and Schultz (2014) showed that ejection angles for loose particulate pumice powder and loose sand were much lower than for a sand block at very early time, however at late times, the difference between pumice powder and loose sand disappeared despite large differences in static material properties. For the purposes of the DART mission, we are most concerned with the moderately early (transient crater) to late time effects, which maybe captured by LICIA Cube, and later by Hera. Further, computational studies at planetary defense scales into asteroid targets suggest that increasing internal friction decreases ejecta angle (Raducan, Davison and Collins 2022). These differences could be due to different results for computations and empirical observations at different time scales. To more fully understand these effects, further investigation is needed into the effects of the internal coefficient of friction of material at planetary defense scales.

Prieur et al. (2017) conducted a study across a range of cohesions (5 Pa — 10 MPa) and coefficients of friction (0.1 — 1.0) to examine proxies for small craters in lunar regolith and evaluate how cohesion and porosity affect crater size. For a porous target (φ=3-20%) with low cohesion (5 Pa), crater diameter decreased by ~10% when the coefficient of friction increased modestly. A smaller crater size resulted as the coefficient of friction increased (fi = 0.8); the crater size decreased by ~30—40% compared to a coefficient of friction of 0.1. Material cohesion (strength) also plays a role. For porous targets with average coefficient of friction (fi = 0.6), crater diameter decreased significantly (~30-40%) as cohesion increased from 0.1 MPa to 10 MPa. These interplays suggest both friction and cohesion are important variables when determining crater size.

It should be noted that this could potentially be an influence of the porosity of the target as well, as pores collapse and absorb energy that reduces the final crater size (e.g., Schultz et al. 2005, Schultz et al. 2007), which illustrates another important parameter we consider in more detail in later sections. In total, previous works showcase that the effects of porosity, cohesion, and internal coefficient of friction all influence the resulting crater size, but are difficult to disentangle since they are dependent on one another in realistic material.

Next, we will briefly discuss the effects of material properties on ejecta behavior in the context of using the observed ejecta dynamics to estimate the material properties of Dimorphos. Here, we rely on experiments and simulations that inform scaling rules that describe crater ejecta (Housen, Schmidt, & Holsapple 1983). The ejection velocity, the total mass of ejecta, and the ejection angle are all observables that can be plugged into scaling rules to inform the initial conditions of the impact event (i.e. projectile properties (which are known), and target material properties (which are unknown), and the gravity field). The DART spacecraft is expected to impact Dimorphos at 6.14 km/s (Design Reference Mission, DART project documents; Rivkin et al. 2021), and LICIACube observations will provide information about the ejecta curtain. Experiments and simulations done prior to impact provide insight into how material properties affect ejecta behavior, and so these observations will provide additional constraints on material properties.

Simulations and experiments show that, like crater size, ejecta behavior variations(*i.e.*, ejection velocity vs. position, and ejection volume vs. time, but not ejection angle) can occur depending on whether formation occurs in the gravity regime or strength regime. However, the effects on each depend on length scales (*i.e.*, the ratio of the launch position of the ejecta, x, and the crater radius, R), where limiting cases (x/R << 1, away from the rim) merge into general cases that have no dependence on strength or gravity (Housen, Schmidt, & Holsapple 1983). Note that ejection angle depends on target material properties, but is independent of regime (strength or gravity, here). In cases where ejecta are affected by material strength and (static) internal friction, impacts into stronger material (and/or with higher friction coefficients) result in a decreased amount of ejecta, lower ejection velocities in a normalized representation, and decreased ejection angles (e.g., Stickle et al. 2015, Luther et al. 2018).

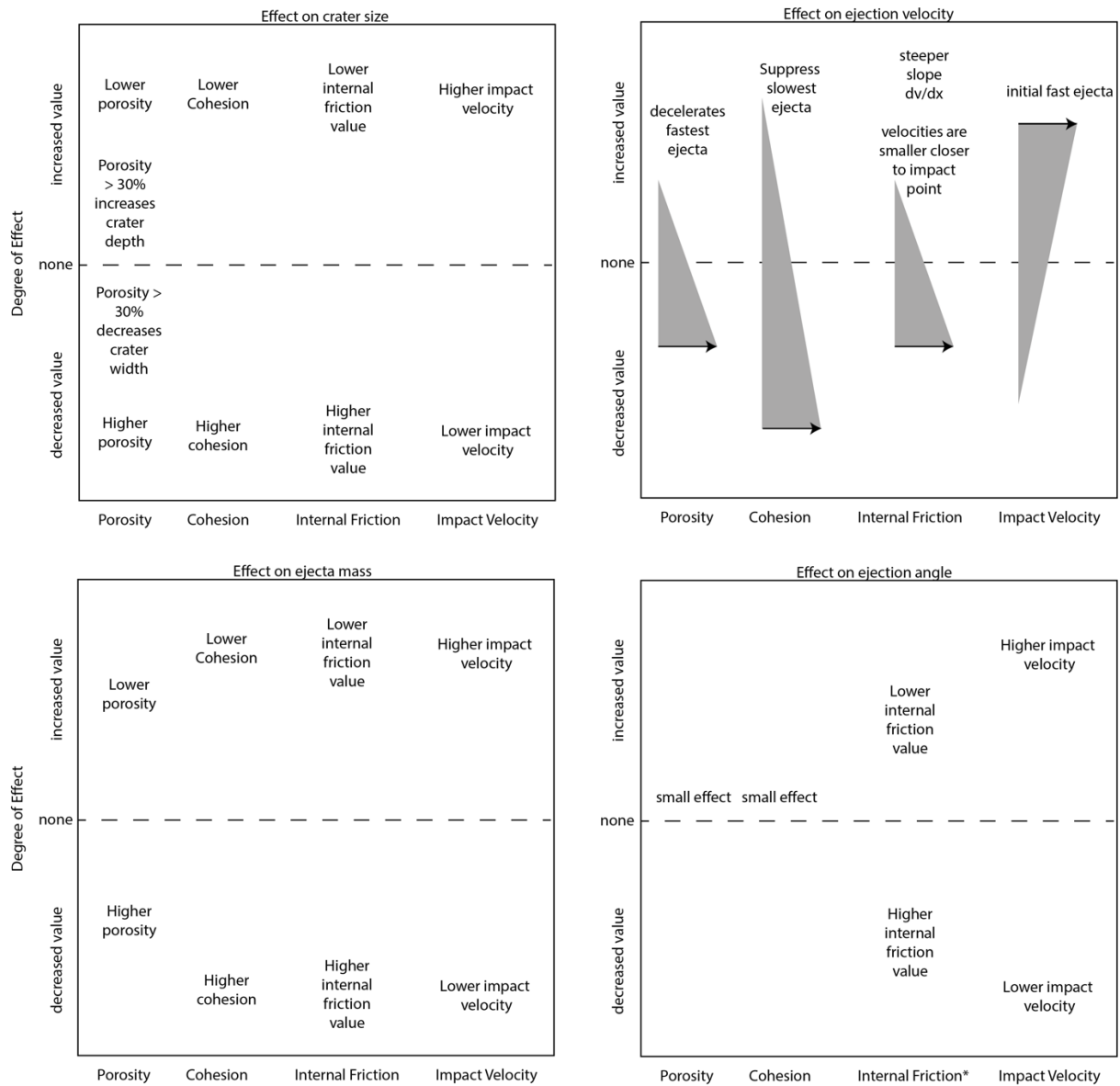

Figure 1. General behavior of the transient crater size (top, left), ejecta mass (bottom, left), ejecta velocity (top, right), and ejection angle (bottom, right) as material properties are varied. Note that these charts are agnostic to gravity and strength regimes as well as dynamic vs. static material properties by showing qualitative trends and assuming conditions before cohesion/internal friction/gravity can play a role at late stages, i.e. the final crater diameter. If no relationship is shown, simulations did not examine the specific interaction in detail. The effects of ejection velocity are shown using triangles: as the value of the property on the x-axis (e.g., porosity,

cohesion, etc.) increases, the triangle shows whether the ejecta velocity tends to increase (base of triangle on top) or decrease (base of triangle on bottom). The descriptions provide additional information. *The effects of internal friction on ejecta angle depends on the specific material and post-shock conditions.

In the strength-regime, the ejection behavior is also affected when the coefficient of friction increases (see Figure 1) (e.g., Hermalyn and Schultz 2014, Luther et al. 2018). A smaller crater size resulting from an increase in the coefficient of friction typically implies a decrease in the amount of ejected material. For this ejecta, simulations show that the ejection angles decrease from an average of ~55° for a coefficient of friction of 0.2 (in a porous target), to ~35° for a coefficient of friction of 0.8 (Luther et al. 2018; See Fig. 1). The lower ejection angles combined with the smaller ejecta mass reduce the momentum imparted in the direction of the incoming spacecraft and reduce β (Fig. 1, 2). Increased cohesion has also been shown to suppresses the amount of the slowest ejecta formed post impact, but did not have noticeable effects on ejection angles in the simulations (Luther et al. 2018).

Luther et al. (2018) used iSALE-2D simulations to examine the expected ejecta curtain for different material properties and impact velocities into solid (non-particulate) targets. Overall, they found that more ejecta are produced for faster impacts. They additionally found that the ejection behavior depends on the individual target properties. This translates into differences in the curtain dynamics and the momentum enhancement. When impact velocity is increased, more ejecta are produced (at constant crater size, i.e., decreasing projectile size) up to the order of the target speed of sound (approximately 4 km/s). Simulations showed that ejection angles increased with impact velocity in the range below the speed of sound, while launch velocities were roughly similar (for absolute positions). As porosity increased, the proximal ejecta (high-speed ejecta with launch positions closest to the impact site) velocities decreased but late-stage ejection angles were relatively consistent across porosity values.

These simulations generally align with results seen in experiments into porous particulate targets from Schultz et al. (2005), who point out that there are two types of porosity that affect crater formation and ejection speeds and angles: bulk target density (porosity) and porosity of the constituent grains (compressibility). Schultz and Gault (1985) showed that even large projectile/target density ratios (>10) do not affect crater scaling for gravity-controlled growth in highly porous targets. However, target density (and, in turn, porosity) does affect peak pressure and its decay rate through the target, directly affecting crater growth (in materials exhibiting some form of strength). Observations from experiments (Schultz et al. 2005) and the Deep Impact mission (Schultz et al. 2007) show that highly porous particulate targets result in two ejecta components (1) a high-angle plume generated during the compression stage; (2) nominal angle related to ejecta flow formed from the outward shock and rarefaction off the free surface. Additionally, a low-density impactor (*b*, the LCROSS experiment) can amplify the high-angle component (Schultz et al 2010). Consequently, observations of the ejecta evolution will be critical in order to understand the coupling of all of these variables, similar to what was done in 2005 to understand the Deep Impact crater. For example, target material that exhibits a high porosity and compressibility may produce a small final crater, which could still be accompanied by a large momentum transfer. Importantly, ejecta curtains will look different as impact velocity and material properties are varied. The full details of the effects of ejecta curtains are well studied and presented in more detail in Fahnestock et al. (2022). Ejecta plume simulations using information from shock physics codes as input conditions show that ejecta particle dynamics are sensitive to initial ejection parameters such as orientation and ejection velocity (Ivanovski et al. 2020). These differences will allow constraints to be placed on Dimorphos's material properties from the ejecta plume observations by LICIACube following the DART impact.

Table 1. Brief summary of the role of static material properties and impact parameter effects on crater formation and ejecta processes for cratering, which may be relevant to the DART impact. Most of these results were drawn from numerical simulations in preparation for the DART mission. This summary is not a complete record of all impact experiments, but is compiled from work done by the AIDA/DART IWG and will be used to interpret the results of the DART impact. For details regarding how the effects were determined, see listed references.

| Target property | Range | Effect on … | Reference |
|---|---|---|---|
| Cohesion, $Y_0$ | 0 — 18 MPa | Crater size decreases with increasing cohesion, shift of gravity/strength regime transition | Stickle et al. 2015 ; Prieur et al. 2017; |
| | | | Caldwell 2019 ; Caldwell et al. 2020 ; Caldwell et al. 2021 (under review) |
| | 0 — 150 MPa | Ejecta: Launch velocity and angle are the same for high cohesion as for low cohesion, but increasing cohesion suppresses the ejecta starting at smallest velocities ("freezing rim") | Luther et al. 2018 ; DeCoster et al 2022 |
| | 0–18 MPa | Ejecta mass and velocity: Analytic models predict less ejecta mass but higher ejection velocities as strength increases | Stickle et al. 2015 |
| Coefficient of internal friction, *f* | 0.0– 1.0 | Crater size decreases with increasing friction | Prieur et al. 2017 |
| | 0.0– 1.0 | Ejecta: Increasing friction reduces the amount of ejecta, reduces ejection velocities in a normalized representation, and decreases ejection angles (i.e., change of ejecta curtain and vertical velocity component) | Luther et al. 2018 |
| | 0% – 50% | Increasing porosity reduces the crater size. | Stickle et al. 2015 ; |

| Porosity, $\Phi$ | | Depth-diameter ratios remain relatively constant for 0-30% and increase for larger $\Phi$, indicating deeper craters | Prieur et al. 2017 ; Caldwell 2019 ; Caldwell et al. 2020 ; Caldwell et al. 2021 (under review) |
|---|---|---|---|
| | 0% - 42% | Ejecta: Increasing $\Phi$ significantly reduces the launch velocity of proximal ejecta (near the impact point), but has nearly no effect on ejection angles | Luther et al. 2018 |
| Impact velocity | 1 — 20 km/s | Ejecta (tested for constant crater size, i.e., projectile size decreases with increasing impact velocity): ejection angles increase with impact velocity in the range below the speed of sound, while launch velocities are roughly similar (for absolute positions); more ejecta is produced increasing impact velocities up to the order of speed of sound → ejecta curtains look different for range of impact velocities | Luther et al. 2018 |

3.3     Deflection Parameter: Momentum Enhancement (β)

There are three main deflection parameters that the investigation team is concerned with: the impact crater, the ejecta dynamics, and the momentum enhancement factor (β). This section describes the effects of target properties (*e.g.*, cohesion, internal friction, porosity) on momentum enhancement, while the previous section discussed the role of target properties on the crater and ejecta. The crater and ejecta dynamics are two physical observables that will help inform the resultant momentum enhancement factor (β) from the impact, where the determination of β is a Level 1 requirement for the DART mission.  In brief, the momentum enhancement factor is the ratio of momentum transferred to the impacted body to the impactor momentum. When the projectile impacts the target at hypervelocity, there are two components responsible for the change to the overall target momentum, the projectile and the ejecta. The component of the momentum contributed by the ejecta (referred to as β-1) amplifies the overall momentum transfer, because it generates additional momentum as it moves in the opposing direction to the target. In simulations, ejecta is defined as material with a void fraction greater than zero but less than one (i.e., fragmented material), with a velocity component that is opposite to the direction of motion of the target, which does not remain attached to the crater rim. Decades of study have indicated that a number of factors contribute to both crater ejecta (mass, behavior) and morphology (size, shape). Historically, the value of β was understood to depend on target material properties and impactor velocity (e.g., Asphaug et al. 1998, Holsapple and Housen 2012, Jutzi and Michel 2014, Stickle et al. 2015). The DART IWG has completed substantial work to expand upon this knowledge. While many of these target properties were varied individually within simulations to systematically study their effects on crater formation and momentum enhancement, note that, in reality, many target properties are not independent of one another and thus have covariances. For example, as porosity varies, so, too, does material strength (this is called "effective strength" in this manuscript). This inter-dependence becomes important for DART as we know little about the material properties and structure of Dimorphos. Therefore, many combinations of material properties may describe the asteroid system as it is today.

Table 2 summarizes the current best understanding of how specific properties may affect momentum enhancement from the DART impact. More details are included in the following sections. In general, simulations suggest that β, is strongly affected by target cohesion (i.e., the shear strength of the material at zero pressure), the coefficient of internal friction, and the target porosity (e.g., Walker and Cochron 2011, Stickle et al. 2017, Rainey et al. 2017, Raducan et al. 2019, Rainey et al. 2020, Raducan and Jutzi 2022, Luther et al. 2022). Other parameters, such as target shape and internal structure, can affect β, but the effects are more subtle and depend on specifics of the shape combined with impact angle and the near-surface structure. These effects are summarized in Table 2, and the general relationships and magnitudes of effects are shown qualitatively in Figure 2.

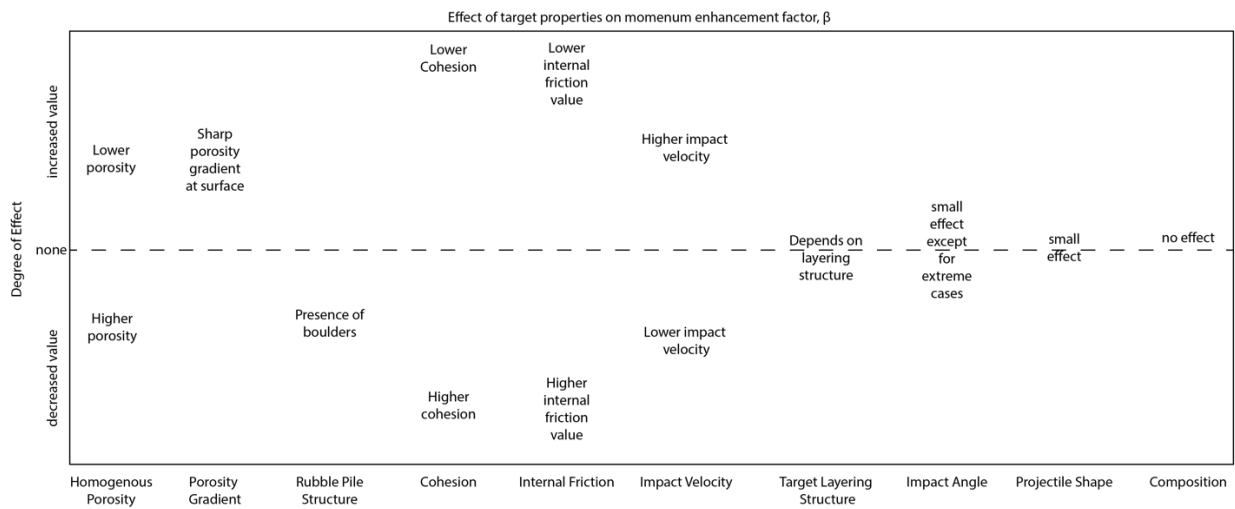

Figure 2. General effects of a variety of target material properties (competent and particulate) on momentum enhancement. The distance from the center horizontal axis (dashed line) is intended to show relative magnitude of the effect.

Table 2. Brief summary of material properties and impact parameter effects on momentum enhancement that may be relevant to the DART impact. This summary is compiled from work done by the AIDA/DART IWG and will be used to interpret the results of the DART impact. For details regarding how the effects were determined, see listed references.

| Target property | Range/Description | Effect on $\beta$ | Reference |
|---|---|---|---|
| Cohesion, $Y_0$ | $Y_0 = 0$ — 100 MPa | $\beta$ strongly increases with decreasing $Y_0$ | Syal et al. (2016); Raducan et |

| | | | al. (2019); Rainey et al. (2020); Raducan and Jutzi (2022); Walker and Chocron (2011); Raducan et al. (2022), Owen et al. (2022) |
|---|---|---|---|
| Effective strength, $Y_{eff}$ (at $\rho$ = 1000-3000 kgm$^{-3}$) | $Y_{eff}$ = 0–300 kPa $\rho$ = 1000 – 3000 kg m$^{-3}$ | $\beta$ increases with increasing effective strength; $\Delta v$ decreases with increasing effective strength | Stickle et al. (2015) |
| Coefficient of internal friction, $f$ | $f$ = 0.4 – 1.2 | $\beta$ strongly increases with decreasing $f$ | Stickle et al. (2017); Raducan et al. (2019); Rainey et al. (2020); Raducan and Jutzi (2022) |
| Initial porosity, $\Phi_0$ | $\Phi_0$ = 0% – 50% | $\beta$ moderately increases with decreasing porosity  Increasing porosity | Walker and Chocron (2011); Hoerth et al. (2015); Flynn et al. (2015, 2017); Syal |
| | | lowers disruption risk | et al. (2016); Stickle et al. (2017); Raducan et al., (2019); Rainey et al. (2020); Luther et al. (2022) |
| Target structure: layering | Weak, porous layer of varying thicknesses overlying stronger, less porous substrate | Layering can cause both amplification and reduction of ejected momentum relative to the homogeneous case | Raducan et al., (2020) |
| Target structure: porosity gradient | Exponentially decreasing porosity with depth with different e-folding depths (note: "e-folding" is the technical term for the distance over which the amplitude of an exponentially varying quantity increase or decreases by a factor of e. ) | An increase in $\beta$ is observed only for sharp decreases in porosity that occur within 6 m of the asteroid surface | Raducan et al., (2020) |
| Target structure: rubble-pile | Different configurations of boulders embedded in cohesionless targets | The presence of boulders reduces $\beta$ compared to a homogeneous target, and | Stickle et al. (2017); Ormö et al., (2022) Raducan and Jutzi (2021); |

|  |  | the magnitude of $\beta$ depends on the initial boulder configuration | Graninger et al. (in prep) |
|---|---|---|---|
| Impact velocity, $U$ | $U = 0.5 - 15$ km s⁻¹ | $\beta$ moderately increases with increasing $U$ | Walker and Chocron (2011); Jutzi and Michel (2014); Syal et al. (2016); Luther et al. (2022) |
| Impact angle, $\theta$ | $\theta = 90 - 30^o$ (from horizontal) | $\beta$ is similar for different impact angles; however, the imparted momentum is reduced as $\theta$ decreases | Stickle et al. (2015); Raducan et al., (2021) Raducan and Jutzi (2022) |
| Projectile shape | Simple projectile geometries with similar surface areas at the point of impact | $\beta$ is only slightly affected. Specific geometry determined whether β increased or decreased. | Walker and Chocron (2011); Raducan et al., (2022); Owen et al. 2022; DeCoster et al. 2022) |
| Impact point with respect to target | Simple projectile impacting at differing | As distance increases, impact angle and crater asymmetry increase | Stickle et al. (2015) |

| center of figure (i.e., center of the illuminated face of the asteroid) | distances from center of figure | Ejected material is concentrated downrange, which lessens the deflection in the orbital velocity direction | |
|---|---|---|---|
| Target composition | Comparing identical impacts into different target material type (composition): granite, basalt, pumice | Composition of rocky material has no significant on β | Stickle et al. (2017) |
| Target shape | 21 different asteroid shapes<br><br>Prolate and oblate ellipsoidal shapes | Spherically shaped asteroids experience little loss in the expected ΔV (as low as 10% across the body), but irregularly shaped asteroids see up to 50% loss in the expected transfer of momentum compared to nominal impacts | Feldhacker et al. 2016; Syal et al. (2016); Raducan & Jutzi (2022) |
| Asteroid spin rate, P | P = 100 s to P = 2.5 hrs | Fast or very fast rotation has no significant effect on β but does | Syal et al. (2016) |

| | | affect disruption risk | |
|---|---|---|---|

3.4     Effects of Target Strength Properties on Deflection Parameters

Numerous studies point to material strength as one the most influential parameters for kinetic impactor deflections (e.g., Syal et al. 2016, Stickle et al. 2017, Rainey et al. 2017, Raducan et al. 2019, Rainey et al. 2020). Indeed, Rainey et al. (2017) found that zero-pressure yield strength (i.e., cohesion) was the most statistically significant predictor for β. These effects have been studied in laboratory experiments and numerical models (e.g., Housen and Holsapple 2011, Walker and Cochron 2011, Holsapple and Housen 2012, Stickle et al. 2015, Syal et al., 2016, Stickle et al. 2017, Raducan et al., 2019, Rainey et al. 2020, Chourey et al. 2020, Luther et al. 2022). In this section, we focus on the results from numerical modeling studies that explored the effects of target strength and frictional properties on crater size, ejecta dynamics, and β. In general, an increase of target strength (e.g., cohesion and/or coefficient of friction) decreases the crater size, the amount of ejected material, and β.

Stickle et al. (2015) used both analytic models and 2- and 3-dimensional CTH simulations to show the effects of material strength on deflection velocity and momentum enhancement following a DART-like impact. For "realistic" asteroid material properties (sand, weak/soft rock (as defined in Holsapple 1993)), the analytic models predicted crater diameters between 8 m and 17 m. Specifically, the models of weak rock, with yield strength of 7.6 MPa, predicted crater diameters of 12 m, while material representing sand, with yield strength of 1 MPa, predicted a crater diameter of 8 m. Material properties equivalent to wet soil produced a crater with a 17-m diameter. For equivalent impact simulations using CTH (90°, through the target center of mass), the crater diameter predicted by simulations is 6 m to 15 m, a difference of 12—25% from the analytic predictions. Predictions for β ranged from ~1 to 3.8 for impacts of differing target strengths, with predicted deflection velocities (*i.e.,* the change in orbital velocity due to the kinetic impactor) of ~0.007 cm/s to 0.147 cm/s (within the DART level 1 requirement of 0.7 mms[-1]) (Rivkin 2021).

Raducan et al. (2019) used the iSALE-2D shock physics code to numerically simulate impacts into low-gravity, strength-dominated asteroid surfaces and to quantify the sensitivity of ejecta properties and momentum transfer to variations in asteroid properties. Dimorphos was represented as a half-space with 2D cylindrical geometry. They found that the cohesion (e.g., Fig. 3) and the internal friction coefficient of the target's damaged material (post-shock) had the greatest influence on momentum transfer, similar to the findings of Rainey et al. (2017). In agreement with Luther et al. (2018), Raducan et al. showed that an increase in target cohesion limited the amount of total ejecta and suppressed the final, slowest, ejecta leaving the crater (e.g., Fig. 1). An increase in the internal friction resulted in lower ejection velocities. Therefore, as the cohesion or coefficient of internal friction was decreased, the $\beta$ increased. Using representative impactor parameters for the DART spacecraft and reasonable estimates for the material properties of the Didymos binary asteroid (e.g., 20% target porosity, DART Project DRA), Raducan et al. (2019) found that $\beta$ ranged from approximately 2.4 for a target with a cohesion of 10 kPa to approximately 4 for a target cohesion of 0.1 kPa. For a target with a much lower cohesion of 100 Pa, the crater size reached up to ~30 m diameter. They also showed that analytically derived β (e.g., Cheng et al. 2017) closely matches numerical simulation results.

The Hayabusa2 Small Carry-on Impactor (SCI) experiment at the rubble-pile asteroid Ryugu, fired an ~300 mm x 300 mm projectile at the asteroid's surface at ~2 km/s and formed a ~15-m-diameter crater. The large diameter suggests that the crater formed in the gravity regime, or in a very low-strength surface (Arakawa et al., 2020). This experiment added to other recent evidence of weak asteroid surfaces found by the Hayabusa2 mission at Ryugu (Arakawa et al., 2020) and the OSIRIS-Rex mission at Bennu (Walsh et al., 2019). Raducan and Jutzi (2022) extended the Raducan et al., (2019) work to study the effects of lower cohesion (down to 0 Pa). The simulation was modeled fully in 3D with the Bern SPH code. Their results, including trends observed in the ejecta mass-velocity distribution from DART-like impacts, were in good agreement with the numerical results from iSALE-2D (Raducan et al., 2019). They showed that impacting a target with fixed 40% porosity and friction coefficient of 0.6, resulted in $\beta$ increasing from ~3.5 for a 50 Pa target to ~ 5 for a cohesionless, 0 Pa (*i.e.,* strengthless) target. When varying the coefficient of internal friction (for fixed target cohesion and porosity, above), an increase from $f = 0.4$ to $f = 1.0$ led to a 25%-33% decrease in $\beta$ (for 50 Pa and 0 Pa cases, respectively). The trends in crater and ejecta behavior with varying target strength observed by Raducan et al. (2019) and Raducan and Jutzi (2022) were in good agreement with results from numerical studies of impacts in the gravity regime (Prieur et al., 2017; Luther et al., 2018). Raducan and Jutzi (2022) also showed that if target cohesion is less than ~10 Pa, the DART impact may produce a structure that is dissimilar to an impact crater (Figure 4).

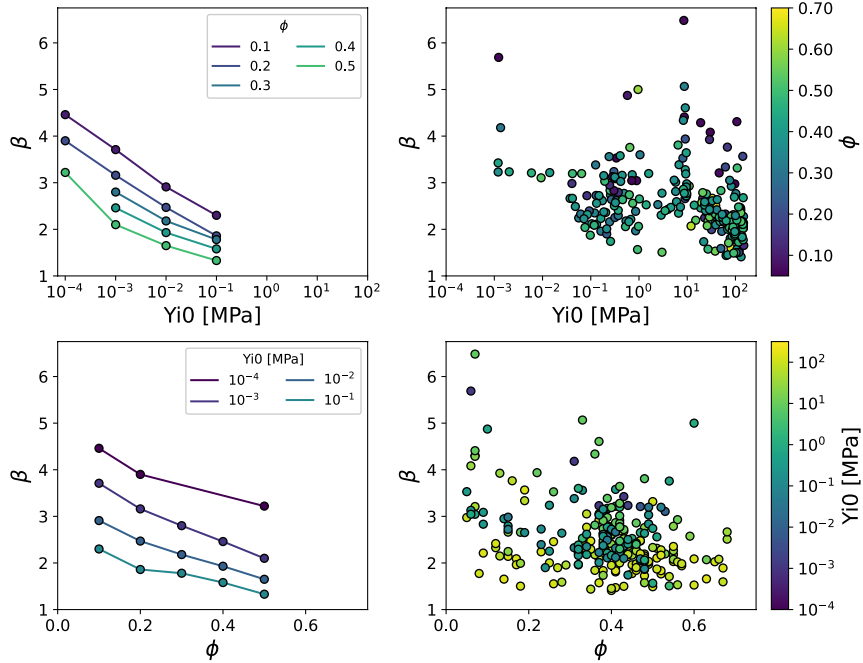

Figure 3. Effects of material cohesion (Yi0) and material porosity (φ) on the momentum enhancement factor (β). (left) Results of simulations showing systematic effects of porosity and yield strength on β (reproduced with permission from Raducan et al. 2019) (right) Results from Spheral simulations, which allowed additional parameters to vary independently of porosity and yield strength. When more than one parameter varied at a time (indicative of realistic materials), immediate trends were not as easily identified. For a detailed discussion of porosity effects, see Section 3.5.

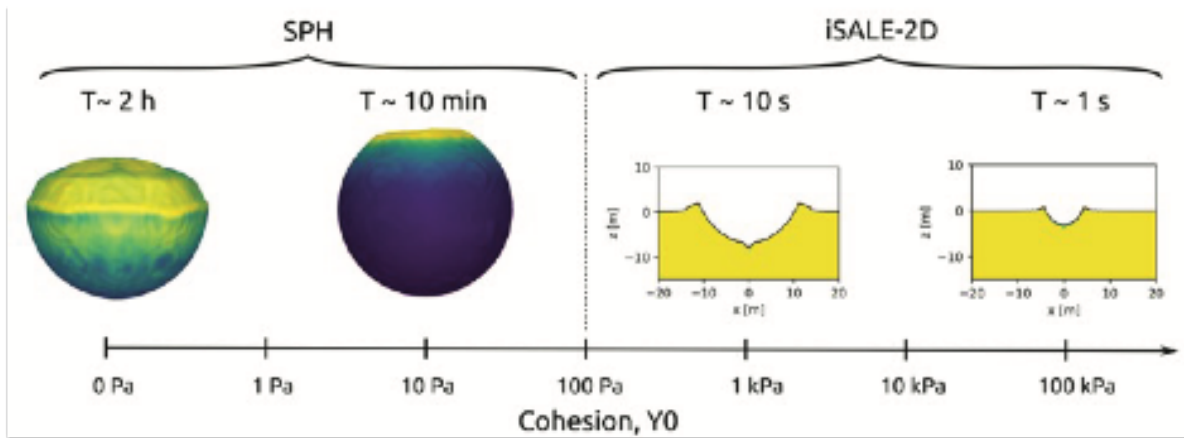

Figure 4. Resulting morphology after a DART-like impact, depending on target cohesion (and fixed 40% porosity). Adapted from Raducan et al., (2019) and Raducan and Jutzi (2022).

3.5     Effects of Target Density/Porosity on Deflection Parameters

This section examines the effects of target density and porosity on deflection parameters of interest for the DART mission. Porosity is understood to dampen the momentum transferred from hypervelocity impacts, as seen both in experimental (e.g., Schultz et al. 2005, Walker and Chochron 2011, Holsapple and Housen 2012, Flynn et al. 2015, Hoerth et al. 2015, Schimmerohn et al. 2019, Chourey et al. 2020) and numerical (e.g., Asphaug et al. 1998, Jutzi and Michel 2014, Syal et al. 2016, Stickle et al. 2017, Raducan et al. 2019, Rainey et al. 2020, Caldwell et al. 2020, Luther et al. 2022) studies. While the porosity of Didymos is estimated to be about 20% (Naidu et al. 2020), the porosity of Dimorphos is still unknown. Studies by Raducan et al. (2019) and Luther et al. (2018) showed that an increase in the initial porosity of the target led to lower ejection velocities (Fig. 1), which in turn led to lower $\beta$ values (Fig. 2 and 4). For fixed cohesion and coefficient of internal friction, a decrease in target porosity from 50% to 10% led to an increase in $\beta$. However, we note that experimental results for pumice powder (porosity of 42%) vs. sand (porosity of 22%) showed little effect of target porosity on ejection velocity, showcasing the major differences between granular and competent material (Hermalyn and Schultz 2014). Further, Hermalyn and Schultz (2011) experimentally investigated the effects of the density ratio (projectile/target) on ejection angles and velocity in porous particulate targets. They found that low-density projectiles coupled closer to the target surface that resulted in lower ejection angles, however a clear relationship between density ratio and ejecta velocity was not observed. Note that these studies did not take into account the effects of different crushing behaviors of the target material (e.g., different crushing pressures, see, e.g., Schultz and Gault 1985, Schultz et al. 2005, Housen et al. 2018). In general, as porosity increased, craters became increasingly narrow and deep, the shape of the subsurface damage region was altered, and, when porosity was very large, ejection angles became high (e.g., Stickle et al. 2017, Caldwell et al. 2020). Additionally, crater profile is also dependent on impact angle, where vertical impacts into highly porous compressible particulates resulted in deep but small transient craters (low cratering efficiency), but efficiency increases with lower impact angles (with respect to the target plane) in the same porous target (Schultz et al. 2005). While $\beta$ decreased with increasing porosity, the deflection velocities for porous asteroids actually increase as a result of the lower total asteroid mass (Stickle et al. 2015, Syal et al. 2016, Stickle et al. 2017). This mass effect outcompeted the loss of kinetic energy to pore crushing and increased $\beta$.

Another simulation component to consider is the modeling of porosity, both the types of porosity and the porosity models themselves. The porosity type could affect the predicted momentum enhancement following a kinetic impact. Microporosity (i.e., porosity included in and between material grains) is the most common method for representing porosity within shock physics codes and is commonly represented using models such as P-$\alpha$ (Hermann 1969) or $\epsilon$-$\alpha$ (Wünnemann et al. 2006). These models consider information about how pores crush out under increasing pressures (commonly called a "crush curve"). Stickle et al. (2020) revealed a noticeable difference (15%—30%) between the momentum enhancement predicted by CTH and Spheral at porosities important for the DART impact (~20%), likely an artifact of the specific porosity model implementations and how they differ in the two codes (see section 2.2 for porosity model definitions). Luther et al. (2022) investigated this further by comparing the resulting momentum enhancement for three codes (iSALE, miluphcuda, Bern SPH). They were able to aligned crush curves between codes and reduce the variation on $\beta$ to +/-5% for most scenarios tested.

Computational studies exploring porosity using CTH showed a decrease in $\beta$ — 1 of 50% when granite porosity was increased from 5% to 20%, though there were no corresponding large porosity experiments (Walker and Chocron 2011). These computational studies also showed that target porosity had a greater influence on $\beta$ than impactor shape when spheres, flat plates, and hollow cylinders were compared. They further showed that target porosity had a greater influence on $\beta$ than impactor density, when spheres of different materials were compared.

The second type of porosity to consider is macroporosity (i.e., fractures or spaces between fragments and boulders within an asteroid). Stickle et al. (2017) showed that when an equivalent amount of microporosity and macroporosity were included in a target, the crater size was much more sensitive to the effects of microporosity, and the amount of ejecta increased. Both deflection velocity ($\Delta V$) and $\beta$ increased for rubble-pile targets (e.g., microporous targets) compared to fully competent or microporous targets with equivalent bulk densities (due to the reduced target mass). Caldwell et al. (2021, under review) also found crater size and shape differences bases on porosity structures, including uniform microporosity and rubble-pile configurations. The effects of rubble-pile structures appear in greater detail in section 3.6 and 3.8.

3.6     Effects of Near-surface and Internal Structure on Deflection
Parameters

In general, the interior structures of asteroids fall into three broad
categories (e.g., Asphaug et al. 2003; Britt et al. 2003): 1) solid objects
made of coherent, strong material ("the monolith"); 2) rubble piles consisting
of a collection of strong pieces within a weaker matrix or held together by
small particles and cohesion (e.g., Itokawa, Bennu, Ryugu); or 3) "fractured
shards" that consist of a solid, relatively strong body that may include large-
scale fractures from previous impact events (e.g., Eros, Ida, Hermione) (Britt
et al. 2003, Marchi et al. 2015 and references therein, Scheeres et al. 2015).
Though we have no direct measurements of the interior structure of asteroids,
formation models suggest that each of these three cases should have different
internal structures, and we can make some assumptions about their general
characteristics.

Recent missions to asteroids showed diversity among asteroid structures in
our solar system. The current scientific consensus is that most solid bodies in
the solar system are more likely to be heterogeneous rubbles piles rather than
homogeneous monolithic bodies. Category 2 (rubble pile) is likely the most
common structure. Rubble-pile asteroids are aggregates held together only by
self-gravity or small cohesive forces (Bagatin et al. 2001, Richardson et al.
2002, Scheeres et al. 2010) and the interior structures may contain a majority
of empty space. Three of the four asteroids that have had spacecraft rendezvous
are rubble piles: Itokawa (Fujiwara et al. 2006; Saito et al. 2006), Bennu
(Lauretta et al. 2019), and Ryugu (Watanabe et al. 2019), with estimated bulk
porosities of ~41% (Fujiwara et al 2006), 25%-50% (Barnouin et al. 2019), and
>50% (Watanabe et al. 2019), respectively. These types of bodies are likely the
results of catastrophic disruptions and re-accretion events. These rubble piles
are likely pervasively fractured and/or have many fragments within a fine-
particle matrix, held together by gravity and cohesive forces. For the binary
Didymos system, Walsh and Johnson (2015) suggest that both asteroids may be
rubble piles. If true, Dimorphos would be unlikely to have a homogeneous surface
structure. In the remainder of this section, we discuss the effects of possible
target structures on the momentum transfer efficiency.

Understanding the effects of subsurface and near-surface structure on the
expected crater, ejecta properties, and momentum enhancement presents a number
of challenges. Not only are the specific strength properties of Dimorphos
unknown, but the total amount of porosity, including the porosity type, is also
unconstrained. While choices can be informed from previous asteroid rendezvous missions (Thomas et al. 2001, Michikami et al 2008, Michikami et al. 2019, Schröder et al. 2021), the boulder size-frequency distribution for rubble-pile asteroids is not well constrained, leading to a large uncertainty in the surface condition of Dimorphos.

Local geology at the impact site affects the cratering process resulting from kinetic impacts. Geologic features may include near-surface layering, porosity gradients, or the presence (or absence) of boulders. For exemple, near-surface layering likely occured on 9P/Tempel 1 based on Stardust NExT observations of a nested crater combined with the ejecta evolution and the total ejecta mass excavated (Schultz et al. (2013)). Raducan et al. (2020) used iSALE-2D to simulate a DART-like impact into two different, non-homogeneous target scenarios in the strength regime: layered targets with a porous weak layer overlying a stronger bedrock and targets with exponentially decreasing porosity with depth. In the two-layer target scenario, the presence of a less porous, stronger lower layer near the asteroid surface led to both amplification of ejected mass and reduction of momentum compared to a homogeneous case. The momentum enhancement changed by up to 90%. On the other hand, impacts into targets with an exponentially decreasing porosity with depth only produced increases in the ejected mass and momentum for sharp decreases in porosity within 6 m of the asteroid surface.

Recent studies (Ormö et al. 2022, Raducan et al. 2021) of impacts into heterogeneous, rubble-pile targets showed that the presence of the boulders within the target led to higher ejection angles with respect to the surface compared to similar impacts into homogeneous targets. Moreover, less mass was ejected and at lower velocities in the rubble-pile targets compared to the homogeneous target scenarios, which resulted in reduced momentum in the ejected material.

These studies (Ormö et al. 2022, Raducan et al. 2021) also considered the size of the boulders relative to spacecraft size. If an impactor spacecraft were to strike a boulder either much smaller or much larger than the spacecraft itself, the cratering process would be unaffected compared to an impact into a homogenous target. The impact surface (e.g., boulder, regolith) determines the amount of ejecta in each case, which affects the amount of deflection. The case for which the boulders were comparable to the spacecraft in size could result in the spacecraft glancing off the boulder or impacting near the edge of the boulder. Such non-ideal impacts could affect the cratering and ejecta processes, which would also change the deflection magnitude. Observations of the asteroid Itokawa show that boulders can affect the cratering process and suggest that boulders can "armor" the surface by fragmenting under impact and reducing the formation of craters (Tatsumi & Sugita 2018).

Two-dimensional simulations (CTH and Spheral) of rubble-pile targets suggested that different boulder-field realizations resulted in a factor of two variation in ejecta momentum enhancement, and the resulting β depended on what the impact site makeup was (boulder, regolith/matrix). Additional 3D CTH simulations showed that for oblique impacts striking a boulder, debris could impact other nearby boulders rather than directly escaping, lowering the value of β. This debris retention is likely an effect of the ejecta being shielded by nearby obstructions, preventing its escape from the target. However, a direct impact on a relatively flat boulder surface without other obstructions could result in a higher β. While 2D simulations are more difficult to draw robust conclusions from due to potential geometry effects, they provide a valuable starting point for building our intuition.

These simulations suggest that boulder configuration can dominate ejecta processes. The size of the boulders, in addition to their spatial arrangement relative to the point of impact, led to cases in which boulders were ejected from the target while their structures remained intact. For rubble-pile simulations, the total β calculated for a variety of boulder-field realizations appeared to scatter within a few to ten percent of the β calculated for the equivalent porous monolithic case.

Graninger et al. (submitted, in revision) used Spheral to simulate impacts into 3D rubble-pile asteroids. In these simulations, an ellipsoid target was filled with boulders ranging in size between 2 m and 20 m. For each simulation, the ellipsoid was rotated slightly about the semi-major axis, which led to the impactor striking into different boulder environments in each case. They found that slight movements, between 40 cm and 100 cm, on the surface of the asteroid resulted in in deflection velocity variations of ~10%. Further, the surface and sub-surface rubble configuration influenced the structure of the resulting impact crater. When impacts occurred centered directly on top of a boulder, the resulting crater was hemispherical with very little asymmetry. If the impact were adjacent to a boulder or if large boulders existed just under the surface, the resulting crater was more asymmetric — either elliptical in shape if boulders were below the surface or having flat edges with offset centers if boulders were on the surface. Caldwell et al. (2021, under review) also achieved a range of crater morphologies for differing boulder arrangements in rubble-pile configurations. In total, while previous simulations provide some insight to the response of rubble pile targets to a kinetic impact event, it is important to note that there is no one-size-fits-all rubble-pile asteroid. This is further showcased by measurements made from various rubble-pile asteroids like Bennu (Barnouin et al. (2019a,b)), Itokawa (Mazrouei et al. (2014)), and Ryugu (Watanabe et al. (2019)), that exhibit vastly different surface characteristics and internal properties.

### 3.7    Effects of Asteroid Shape on Deflection Parameters

Local topography related to an asteroid's shape has been shown to significantly influence the efficiency of deflection by kinetic impactor. Material ejected perpendicular to the surface plane at the impact location could affect the momentum transfer in the orbital velocity plane. Feldhacker et al. (2016) and Syal et al. (2016) examined the effects of irregular asteroid shapes on the efficiency of kinetic impactors. Using radar-derived shape models for a collection of 21 asteroids, Feldhacker et al. (2016) showed that asteroid topography has a significant influence on the momentum transfer to an asteroid following a kinetic impact. In the case of the asteroid 6489 Golevka (which has a particularly complicated shape), losses of up to 34% in the expected $\Delta v$ occurred for some nominal impact locations compared to what would be expected if all $\Delta v$ occurred in the direction of interest (i.e., the asteroid velocity direction). The study also showed that the effective $\Delta v$ imparted to an asteroid varied not only with local topography of an asteroid but also with the asteroid's overall shape. More spherically shaped asteroids experienced little loss in the expected $\Delta v$ (as low as 10% across the body), whereas irregularly shaped asteroids saw up to 50% loss in the expected transfer of momentum, depending on impact site. DRACO observations in the minutes leading up to the DART impact, combined with LICIAcube images, will be used to construct a shape model for Dimorphos, allowing simulations following impact to include appropriate topography and structure around the impact site. The use of such a shape model is expected to minimize the errors and uncertainties associated with topography and shape effects.

Off-axis impacts, which arise from imperfect targeting as well as influences of asteroid shape and rotational state, represent an additional source of variability in deflection response. This topic is discussed in the next section.

### 3.8    Effects of Impact Angle and Local Topography on Deflection Parameters

The effects of impact location, and therefore local slope, on crater geometry have been studied in detail, but the specific effects on the deflection efficiency parameters of interest from a kinetic impact remain less clear. The majority of investigations of hypervelocity impacts for planetary defense consider trajectories of projectiles that are normal to the surface of the target. However, both oblique and off-center impacts occurring between the spacecraft and the asteroid must be explored.

The DART impact is set to occur at an impact angle that depends on both the spacecraft's incoming trajectory and on the local slope of the target at the impact point (Pajola et al., under review). DART will approach Dimorphos at an angle of ~10° from the orbital plane, but little is currently known about the local topography at the impact site. The idealized scenario of a vertical impact through a spherical asteroid's center of mass (or, in DART's case, the center of figure of Dimorphos), in the direction of intended deflection, represents a best-case scenario for maximizing the effective $\beta$. Any deviation from this scenario would result in an oblique impact, decreasing the delivered $\Delta v$ as an asymmetric shock wave (e.g., Dahl and Schultz 2011) and increased ejecta momentum would be directed downrange (Anderson et al. 2003, Anderson et al. 2004). The effect of such deviations depends on local slopes. For example, an oblique impact relative to the overall shape of the target may not be oblique relative to the local slopes. In addition, an impact angle below 20° (with respect to the local target plane) on local slope (e.g., wall of a subdued crater) could result in an increase in $\beta$ due to subsequent impact by surviving pieces of the projectile downrange striking locally at a higher angle and creating an increase in ejecta mass. For example, experiments reveal that impactor debris from a 15° impact could carry away 30% of its original energy (Schultz and Gault (1990)). When the projectile debris hits a downrange facing slope, the resulting crater could be much larger than the primary impact crater (Schultz and Wrobel (2012)), thereby contributing more reverse ejecta. Schonberg and Taylor (1989) also showed that as impact obliquity increased on aluminum targets, the damage morphology of the target changed notably, resulting in debris clouds containing relatively large, fast-moving, ricocheting particles. Unless the simulations fully capture the impactor, this component may not be detected. Further, the effects of oblique impacts are showcased in Figure 5. In each case shown, the impact was "vertical" in the frame (i.e., the impact velocity vector was in the negative z direction); however, the local slope at the impact point resulted in an increasingly oblique impact. For an impact without obliquity (normal impact, 90°, Figure 5A), the crater was symmetric, and the ejecta distribution was symmetric about the impact point. This scenario provided the maximum deflection along the vertical direction. For slightly oblique impacts (Figure 5B, 21° with respect to surface normal), the ejecta was still distributed symmetrically about the surface normal but was no longer parallel to the vertical direction. Thus, the ejecta direction resulted in decreased deflection in the vertical direction because some deflection occurred in the horizontal direction. For an "oblique" impact (here, > 45°), an asymmetric ejecta distribution is seen (note the gray material heading downrange), which further decreased the ejecta contribution to deflection in vertical direction. For an oblique impact of 66° (Figure 5D), some ejecta and projectile momentum were carried downrange of the impact and did not contribute to deflection. The momentum transferred in this case was less than the incoming projectile momentum. Similar behavior was seen in Syal et al. (2016), who also found that off-axis impacts could induce perturbations to an asteroid's rotation and could lessen the linear momentum transferred by the impact.

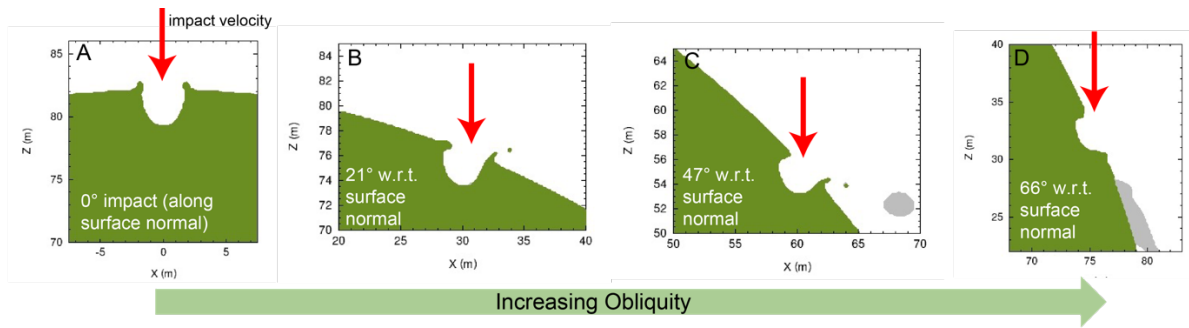

Figure 5. CTH simulations illustrating the effects of increasing obliquity on ejecta direction following a hypervelocity impact. As the impact angle with respect to the surface normal increased (i.e., an increasingly oblique impact), the ejecta and projectile (gray material) momentum was carried downrange and, in some cases, did not contribute to the deflection.

CTH simulations by Stickle et al. (2015) showed that models for a fully competent, strong (yield strength of 200 MPa) asteroid material generally agreed with predictions from analytic studies. For fully dense, competent, strong rocks, values for β were found to be between approximately 3.8 and 5.5, with an imparted Δv to the moonlet of 1.47—2.29 mm/s for impacts hitting at various distances from the target asteroid's Center of Figure (COF) (0—30 m). These β values compare favorably with the analytic solutions based on scaling laws (Holsapple and Housen 2012, Housen et al. 1983, Richardson et al. 2007, Housen and Holsappe 2011) for competent rock, indicating that the impact location with respect to the COF would affect the momentum transfer in the direction of interest (i.e., the orbital velocity direction of Dimorphos).

Raducan at al. (2021) performed iSALE-3D simulations of DART-like impacts on asteroid surfaces at different impact angles and found that the vertical momentum transfer efficiency ($\beta$) was similar for different impact angles; however, the imparted momentum was reduced as the impact angle decreased (with respect to the surface). The expected momentum imparted from a 45° impact could be reduced by up to 50% compared to a vertical impact. In this case, the direction of the ejected momentum would not be normal to the surface, and this quantity would determine the $\varepsilon$ in *Eq.2*. As the crater grew, however, the ejected momentum was observed to "straighten up" relative to the surface (i.e., increased in the positive z direction).

Similar studies were conducted in Spheral using a plate impactor that impacted in one constant direction (e.g., −z) at various distances from the center of an ellipsoid, which resulted in varying impact angles. To capture the effect of changing the regional slope of the target, the impactor angle was held constant and was thus determined by the regional slope of the ellipsoid. For both rubble-pile targets and homogeneous targets, as the regional slope increased, the total momentum enhancement decreased. This relationship follows from the velocity components of the resulting impact ejecta plume; at higher inclinations, not all of the ejecta was normal to the surface. This study also found that the effect of regional slopes on the imparted momentum was greater than the influence of local topography (e.g., boulders; Fig 6).

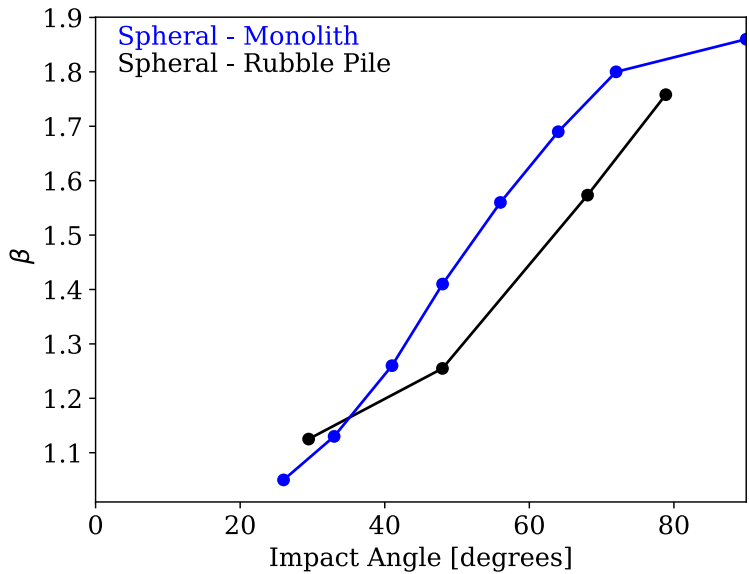

Figure 6. The momentum enhancement in the orbital direction (parallel to impact momentum vector) (β) was affected by the impact angle (measured from the plane that is tangent to the local target surface at impact). Spheral simulations suggested that impact angle had a greater effect on β than near-surface structure and boulder realizations.

In the case of the DART impact, the DRACO camera is set to autonomously navigate the spacecraft to impact Dimorphos as close to the center of figure as possible. However, in reality, DART will likely result in a slightly off-center impact with Dimorphos, and the local surface topography at the impact site could introduce a large impact angle (with respect to the surface tangent). Therefore, we investigated the sensitivity of the imparted period change as a function of offset from the center of figure and impact angle. In general, we find that as impact angle increased, the total change in momentum decreased. An impact angle of 70 degrees (measured from normal) seemed to be a critical angle for which the change in ejecta momentum and the overall change in the period of the asteroid dropped tremendously.

### 3.9     The Effects of Impact Velocity on Deflection Parameters

The DART spacecraft's impact velocity depends on the time of impact and will be a known quantity. In order to fully understand the potential deflection results and how different impact parameters affect β, though, it is important understand the effects of impact velocity. Impact velocity strongly affects the cratering process and influences the deflection response following a kinetic impact.

Holsapple and Housen (2012) used laboratory experiments to derive point-source scaling relationships and determined that the momentum carried away by the ejecta $(\beta - 1)$, in the vertical direction scales with the impact velocity (U~ 6 km/s) as:

$$\beta - 1 \sim U^{3\mu-1} \quad (3),$$

with $\mu$ representing the velocity exponent and taking values between ⅓ and ⅔ (Schmidt 1980, Housen and Holsapple 2011). Numerical studies allow these relationships to be extrapolated to impact velocities beyond the range of typical laboratory experiments. Jutzi and Michel (2014) studied the effects of impact velocity on $\beta$. They used the Bern SPH code to model impacts into strong basalt targets at impact velocities between 0.5 and 15 km/s. They found that $\beta$ increased with increasing velocity, as predicted by the Holsapple and Housen (2012) scaling. However, they found that the velocity exponent $\mu$ exceeded theoretical predictions. Syal et al. (2016) also achieved simulation results consistent with power-law descriptions by Holsapple and Housen (2012) by simulating impacts ranging from 1—30 km/s, but they noted that the velocity scaling component, $\mu$, varied depending on target type.

3.10    Effects of Projectile Mass and Shape on Deflection Parameters

For simplicity, most numerical simulations of the DART impact assume that the projectile is an aluminum sphere, which allows for axial symmetric simulations and reduces the need for resolving thin-walled structures (e.g., a spacecraft). The differences in resolution requirements for a small, thin-walled spacecraft structure compared to a much larger asteroid (e.g., Fig. 7) can be extremely computationally expensive. However, the DART spacecraft is significantly different than a compact sphere, with an under-dense main spacecraft bus and long solar-panels (e.g., Figs 7, 8), which could affect the cratering process and resulting momentum enhancement. For example, experiments done in preparation for the LCROSS impact suggested that hollow and under-dense projectiles could create abnormal ejecta patterns and generate high-angle plumes (e.g., Schultz et al. 2010, Hermalyn et al. 2012). Additionally, experimental investigations into the effect of impactor density on cratering efficiency showed that dense projectile penetrate deeper and couple later in the target, whereas under-dense projectile couple quickly and much closer to the surface (Hermalyn and Schultz 2011). These differences lead to non-proportional crater growth where craters formed from low density projectiles exhibit outward growth due to their shallow coupling depth, and craters from high density projectiles exhibit more downward growth before transitioning to outward expansion. Hermalyn and Schultz point out that the effect of projectile density on the depth of coupling has implications for the source depth of ejecta from primary craters. Therefore, in order to better understand the necessity of simulating the DART impact using a complete spacecraft model versus a simplified model, the effects of different projectile geometries were investigated in several studies.

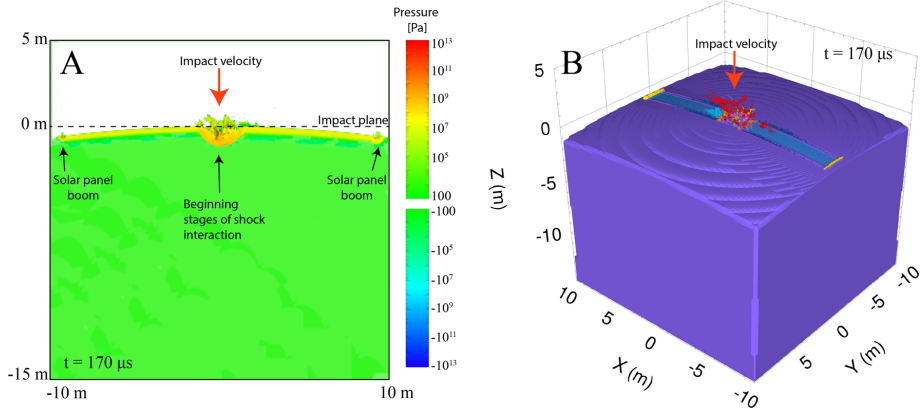

Figure 7. CTH simulations of the complete DART spacecraft model impacting an asteroid at ~6 km/s. A) Shock wave at 170 μsec after impact, showing the elongated region of interaction from the large solar arrays. The strongest shock wave was generated by the spacecraft bus itself. B) Representation of materials 170 μsec after impact, showing the solar panels reaching across the surface and the fragmenting spacecraft.

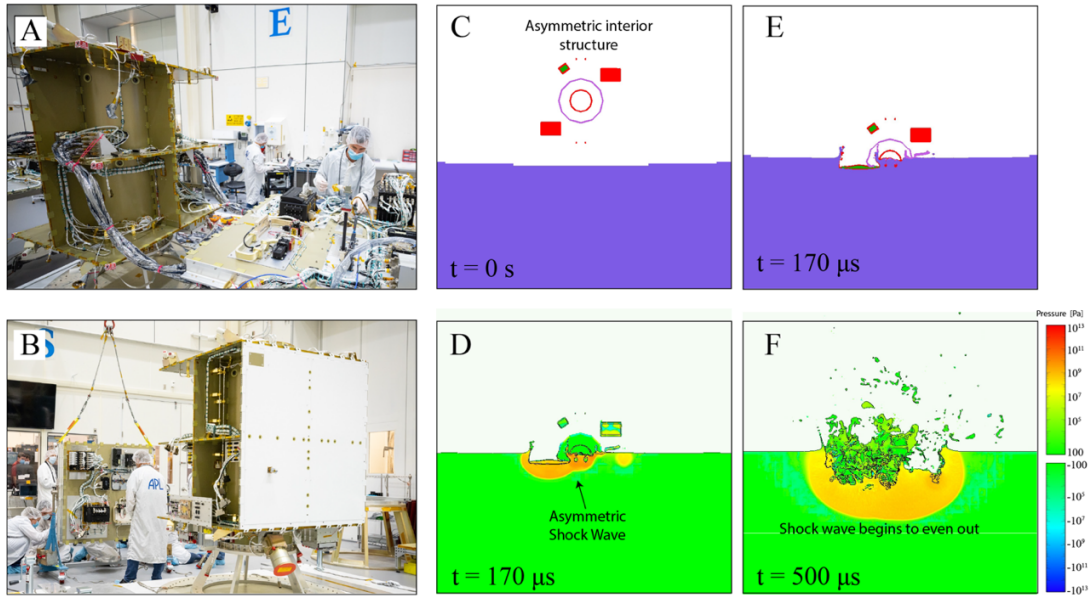

Figure 8. A,B) Pictures of the DART spacecraft during assembly. Note the under-dense main spacecraft bus and regions of higher mass resulting from the variety of components. A) DART panel integration and testing, July 2020. Photo credit: JHUAPL. B) DART Panel Closeout, September 2020. Photo credit: JHUAPL. C) CTH simulation of a spacecraft model, including the main massive components of the interior structure. The simulation was performed in 3D, and the figure is from a 2D slice through the center plane of the spacecraft. D) Pressure in the asteroid and spacecraft 170 $\mu$sec following impact, showing development of an asymmetric shock wave as different components hit at different times. Negative pressure represents tension, and positive pressure represents compression; E) Representation of materials 170 $\mu$sec after impact, showing how massive components hit at different times. F) Shock wave at 500 $\mu$sec after impact. Despite the initial asymmetries, the shock wave begins to regularize at later times. Negative pressure represents tension, and positive pressure represents compression. The viewing window of figures C–F is 2 m x 2 m.

Simulations also showed that asymmetrical interior spacecraft structure affected crater evolution and coupling of the spacecraft to the target during impact. These observations are key factors in the time immediately following impact (Fig. 8). As the shock wave evolved, however, it tended to regularize, reducing the effect of the internal spacecraft structure. It should also be noted that even though the shock wave evens out, the evolution of ejecta may vary. Experiments from Schultz and Gault (1985) using clustered impacts revealed a more chaotic distribution of ejecta until very late stages of the ejecta plume evolution, consistent with inferences drawn from Fig. 8. However, the late stage ejection represents a minor component. Consequently, even though the shock wave regularized, the potential effect will be to reduce $\beta$. Moreover, these results suggest that the momentum enhancement factor is affected more by the velocity distribution (speed and angle) of the ejecta, rather than the total ejecta mass.

Raducan et al. (2022) investigated the effects of simple projectile geometries on the DART impact outcome using the iSALE shock physics code in 2D and 3D. They found that simple projectile geometries with similar surface areas at the point of impact had minimal effects on the crater morphology and momentum enhancement. The crater radius and the crater volume were affected by less than 5%, while the effects on the momentum enhancement was within 7%. In the case of a more extreme projectile geometry (i.e., a rod, modeled in 3D), the crater was elliptical and 50% shallower compared to the crater produced by a spherical projectile of the same momentum. In this case, $\beta$ was within 10% of the simple case. Additionally, Schultz and Gault (1985) suggested that for hypervelocity impacts occurring at small spatial scales (i.e. lab scales), the transfer time of the momentum from impact to target is small (10 $\mu$s) compared to the time for crater formation (10–100ms), so that late stage growth consumes the signatures of penetration. However, as the time for momentum transfer comprises more of the crater growth at larger scales (i.e., planetary defense scales like DART), a projectile geometry effect that controls the penetration phase becomes more evident and may persist to later times.

Owen et al. (2022) investigated projectile geometry effects by comparing simple impactor shapes with constant masses (including a single solid sphere, multiple solid spheres, and cylinders with varying aspect ratios) to a high-fidelity model representation of the DART spacecraft using three different codes: Spheral, CTH, and iSALE. They found that in all cases, a simple spherical projectile overestimated both the crater size and the momentum enhancement factor ($\beta$) compared to the more complex DART spacecraft model. Although the sphere impactor resulted in a larger crater and more ejecta mass than the model of the actual DART spacecraft, simulations showed that the enhanced β stemmed from an increase in the early-time ejecta velocity rather than the increased amount of ejecta mass. These findings are consistent with results found by Raducan et al. (2022). Owen et al. also modeled impactors elongated along the direction of impact (Figure 9): DART spacecraft rotated so that a solar panel wing hits before the main bus, followed by the second solar panel wing; three spheres arranged in series; and the tallest/narrowest cylinder (Diameter=50 cm). These impactor models demonstrated some shielding/capture of the ejecta in the crater volume from interactions with the narrower crater walls, which inhibited the production of ejecta. In agreement with previous studies (Raducan et al. 2019; Stickle et al. 2020. DeCoster et al 2020), Owen et al. showed that β, ejecta distribution, and the crater morphology were more sensitive to target material strength than projectile geometry. They reported variations in ejecta mass ranging from ~10 x mass of DART to ~1000 x mass of DART for strong to weak targets, respectively. When the target's solid yield strength (cohesion) was about 100 MPa, spherical impactors overestimated β by 5—15%, while in a weaker target with solid yield strength of about 0.1 MPa, the overestimation was 10—20%. In total, Owen et al. (2022) note that this geometry effect is minor compared to other variables within the impact scenario, such as material strength properties, impact angles, local surface topography, and porosity, all of which can affect ejecta momentum by more than 200%.

Of the idealized impactors considered by Owen et al. (2022, this issue), the use of three separate spherical projectiles aligned in a plane (to mimic the solar panel wing booms and main spacecraft bus) provided the best approximation for the crater size, ejecta distribution, and β. This approach was shown to be better than attempting to optimize the size of the impact surface (through changing the aspect ratio of cylinders) to best match the DART spacecraft. Overall, these results suggested that there are projectile geometry effects that should be considered when modeling the DART spacecraft. Modeling DART as a single sphere represented a limiting case that over-predicted β by 5-20% and over-predicted the crater size by 79-147% (assuming bowl shaped craters). In simulations, the DART spacecraft behaved more similarly to multiple impactors than to a distorted single projectile (Fig 9).

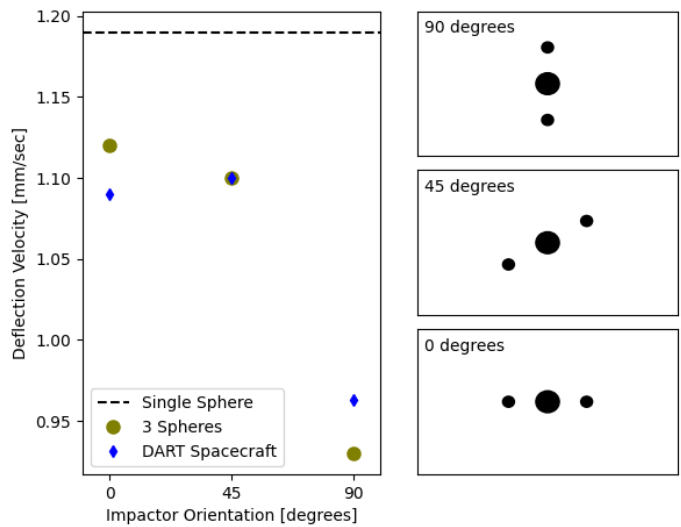

Figure 9. Studies of the effects of projectile geometry suggested that using 3 offset spheres was a better representation of the DART impact than a single spherical projectile of equivalent mass. (left) results for Spheral simulations showing deflection velocity (for the case of a relatively strong Dimorphos) as a function of the orientation angle of the impactor (shown on the right). The orientation angle is measured from the plane that is tangent to the target surface, where 90° details impactors in series and 0° details impactors aligned parallel to the target surface. A three-sphere representation was a much closer match to the predicted deflection velocity from the full spacecraft model than a single sphere impactor.

### 3.11    Experimental Constraints on β

The DART impact modeling working group uses all available methods to study and understand the effects of various material properties on momentum enhancement from a kinetic impactor. This includes numerical and experimental studies.   Here, we briefly summarize some results from relevant recent experimental studies that can provide some intuition for better understanding the DART impact.

Over the years, momentum enhancement (β) has been measured in a variety of experiments. Two important impact parameters come to the forefront in these studies: impactor size and target porosity. Because β increases with increasing impactor size and decreases with increasing target porosity (neglecting mass differences between porous and non-porous targets at large scales), these characteristics compete with one another to affect the momentum enhancement.

First, we examine the effects of projectile size. In experiments into solid materials, $\beta$ increases as the impactor size increases. This general trend was noted in the 1960s for aluminum targets (Denardo and Nysmith 1964), with more recent experiments showing the trend for larger impactors (~3-4.45 cm) into granite (Walker et al. 2013, Walker et al. 2020), sandstone (Schimmerohn et al. 2019) and in concrete (Chocron et al. 2019). In total, these experiments revealed a significant size effect on momentum enhancement where $\beta$ −1 increased as $(D/Dscale)^a$, where $D$ is the projectile diameter, $Dscale$ is a normalization factor, and the exponent $a$ ranges from 0.4-0.65 (Walker et al. 2013, Walker et al. 2020, Chocron et al. 2019). For projectiles larger than ~4 cm, there are insufficient data to determine whether this size-scaling behavior continues or if it saturates. In the aluminum target tests reported in Walker et al. 2020, the ejecta mass saturated while $\beta$ did not, which may be a phenomena that occurs at larger planetary defense scales, like DART. A recent test of a 3-cm-diameter aluminum sphere at 5.4 km/s into an assembly of small rocks (with sizes roughly that of the impactor, representing a rubble pile) held in place with concrete showed $\beta$ = 3.4 (Walker et al. 2022), which is significantly higher than what was seen for the solid target cases. This test result is likely a lower bound, as the target failed at the sides, allowing material to exit laterally.

The second parameter under consideration in the experiments described here is the role of target density and porosity ($\varphi$). Solid pumice targets ($\varphi = 70$-$75\%$, $\rho = 0.7$-$1\,\mathrm{g\,cm}^{-3}$) were impacted with 2.54 and 4.45 cm diameter aluminum spheres at 2.1 km/s (Walker et al. 2017). The resulting $\beta$ values ranged from 1.0 to 1.7. While the direct effects of porosity on $\beta$ were not determined, the resulting $\beta$ is consistent with scaling rules (*e.g.*, Holsapple and Housen 2012) and computational work (*e.g.*, Jutzi and Michel 2014) that indicate that impacts into highly porous ($\varphi \geq 50\%$) targets result in a small $\beta$. Given the lack of large impactor data, we mention work performed with 0.5-cm-diameter impactors into porous rocks (porosity ranging from 25-87%). These experiments showed $\beta$ of 1.75 to 2.25 for impact speeds of 6 km/s (Hoerth et al, 2015) and 0.3175-cm diameter impactors into pumice, which gave $\beta$ values of 2 to 2.5 for impact speeds of 4 km/s (Flynn et al. 2015). Thus, we expect porosity to reduce the momentum enhancement. It should be noted, however, that the pumice tests should represent lower bounds and, given a lower porosity for Dimorphos than for pumice, $\beta$ could be larger than seen for pumice.

All in all, a speculative prediction for the momentum enhancement of the upcoming DART impact based on extrapolation of these experiments (and considering size and velocity effects) indicates a β of at least 3, perhaps much larger.

3.12 Understanding the Ejecta Offset Direction

Equation 2 (Section 1.2) represents an exact result for β for a general impact geometry. Within equation 2, recall that the small vector $\vec{\epsilon}$ is an offset vector between the surface normal and the ejecta direction. While the evolution of the ejecta velocity vector with time is complicated, the post-impact "average" defined by $\vec{\epsilon}$ is straightforward to pull from any modeling result in which the total ejecta momentum vector and/or the post-impact deflection vector were calculated. The DART project adopts the standard 1-dimensional, normal-component enhancement factor for β, where the incoming spacecraft impacts the asteroid dead-center, giving it a push in the spacecraft's direction of motion, enhanced by the additional momentum imparted from ejecta that is directed back along the spacecraft's path (which is assumed to be normal to the target) (Rivkin et al. 2021). The definition of β along the surface normal requires the distinction of 3-dimensional (3D) components of ejecta velocity, and deflection velocity, to ensure only the normal component contributes to the reported β. Although this is conventional within the impact community, the situation is 3D. In order to account for this, we report the parameter epsilon ( $\vec{\epsilon}$ ) (Equation 2 Section 1.2), which expresses the component of non-normal ejecta momentum. Except for extreme cases (e.g., very oblique impacts), $\vec{\epsilon}$ should be small (e.g., Fig. 10) and can be determined by comparing the ejecta direction to the surface normal from any impact modeling simulation. This offset is a result of the impact geometry and does not affect the way impact simulations are set up or analyzed for the DART project. However, it can provide some insight into the cratering process and potentially provide additional constraints on Dimorphos's material properties. Calculations of $\vec{\epsilon}$ for a set of CTH and Spheral calculations, all in the strength regime, show that $\vec{\epsilon}$ can vary based on material properties (Fig. 10), which is expected because ejecta dynamics are also a function of material properties.

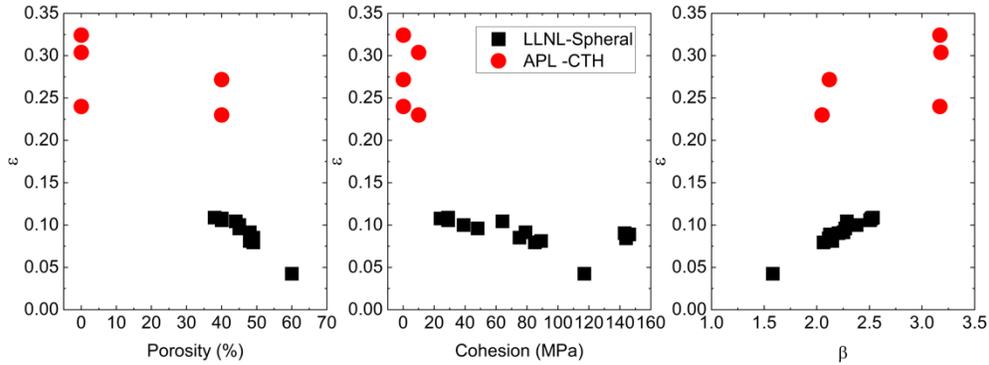

Figure 10: Resulting ε from hypervelocity impact simulations into solid basalt targets, the offset vector of ejecta from the surface normal direction, can be affected by impact geometry and material properties of Dimorphos. Though expected to be small, this offset can provide information about the porosity (left) and material strength (middle) of Dimorphos. (right) The momentum enhancement factor (β) is also affected by the angle at which material is ejected from the surface. Note these represent offsets of only ~6-10° for Spheral calculations and up to ~20° for CTH calculations shown here.

## 4  IMPLICATIONS FOR UNDERSTANDING THE DART IMPACT

### 4.1  Lessons Learned from Deep Impact

Like the combination of DART and LICIACube, the Deep Impact (DI) experiment included an impactor and a follow-on spacecraft to image the ejecta. These images provided vital information about the material properties of Comet Tempel-1. Experiments provided important information about how material properties affect the ejecta curtain to interpret those images; for DART, we will use data from experiments and simulations described in the present work. For example, a low-angle, opaque ejecta cloud at early times indicates a rapidly coupled impact (e.g., there is not much penetration below the surface, as might be expected for a strong target), while a high-angle ejecta plume (at early stages) — or even the presence of a plume uprange — suggests deeper coupling over longer time frames, which would point towards a highly porous target (Schultz et al. 2007). Experiments and simulations into very weak and porous solid targets show that as porosity increases the ejection angle for material steepens (see section 3.2). This relationship was also seen in DI images. The angle of the ejecta can thus provide first-order estimates of porosity and inform the initial conditions for impact simulations. It is important to note that pre-mission predictions for DI suggested that the effect of porosity only appears when porosity becomes extreme (>60%), indicating the lack of a systematic progression between cratering and ejection angles as a function of porosity observed for particulate targets (Schultz et al. 2005). The DI impact crater was controlled by both strength and gravity, however the ejecta curtain indicated gravity-controlled growth, including disruption of the ejecta curtain by pre-existing topography (Schultz et al. 2005). Stereo images from Stardust-NExT revealed nested craters indicating that target material consisted of loose, particulate surface layers on top of a more competent substrate (Schultz et al. 2013). Therefore, observing the ejecta provides partial clues to target morphology, however pre- and post-impact imaging are important evidence for understanding the full picture. Shock asymmetries and energy losses resulting from compaction, comminution, and/or pore crushing are manifested in the distribution of ejecta velocities (direction and speed). As target porosity increases, asymmetry in ejecta flow field persists to later times (Schultz et al. 2007); should the DART impact occur in the gravity regime, however, this asymmetry should be much less pronounced. In the case of DI, the ejecta cone remained attached to the surface, which indicated that formation of the crater was controlled by gravity rather than by strength (A'Hearn 2005). Indeed, estimates for the shear strength of the material were ~65 Pa (A'Hearn 2005), orders of magnitude lower than many targets in strength-dominated impacts. Evaluation of the ejecta cone behavior following the DART impact could help identify the crater processes that are dominant on Dimorphos. Images from the trailing spacecraft of DI also showed significant asymmetry in the ejecta cloud at early times (which extended to late times) as a result of the oblique impact. Evaluating the ejecta cone following the DART impact, including how much it is offset from the expected surface normal direction (i.e., $\varepsilon$), will provide additional constraints on the impact angle and the local geology at the impact site.

4.2    Understanding Beta and the Correlation to Material Properties

The studies presented in previous section found that similar deflections (i.e., similar $\beta$ values) can be achieved by impacts on targets with very different material properties or near-surface structures. The interaction of material parameters can be examined in parallel coordinate plots (Fig. 11, Fig. 12). These plots showcase the complicated relationship between target material properties (strength and porosity) and the momentum enhancement factor. In these plots, each line represents a single simulation, where the relevant input parameters can be traced along a given line.

Figure 11 examines a small number of simulations from a single code (iSALE) to illustrate the utility of these plots and point out some initial important relationships. The color bar gradient (representing β) is lighter (higher β) for simulations with initial conditions that had decreasing cohesion, porosity, and coefficient of internal friction. The clearest trends are given by the variables that have the strongest effects on β. For example, friction and cohesion dominate over porosity. For this set of simulations, the cohesion values are plotted on both sides of the figure to allow correlation between cohesion and porosity, between porosity and friction, and between cohesion and friction.

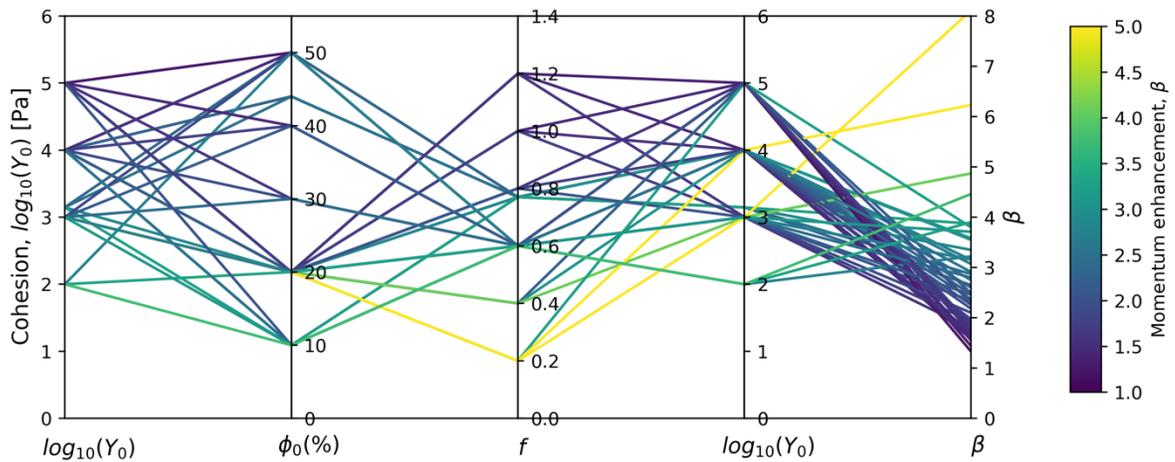

Figure 11. Parallel-line plot from a subset of iSALE simulations showing the relationship between porosity (φ), coefficient of friction (fi) and material strength/cohesion (Y0) and the resulting momentum enhancement. The lines are colored by the value of β. A range of material property combinations can produce similar momentum enhancement.

These relationships become more complicated when additional material parameters, and additional codes, are examined. A larger suite of simulation results, from a variety of simulations across the IWG, is shown in Figure 12. Because of the number of simulations performed by the IWG, many lines overlap, especially on the left-hand axes. In that case, simulations resulting in lower β values are plotted as thicker lines under simulations predicting higher β values. The lines are colored by the β value predicted by the simulation. Note that the density of lines for a given range in results should not be interpreted as the probability of deriving a certain momentum enhancement factor from the DART impact. The line density merely represents the sample parameter choices in the simulations, and may be affected by sparsely sampled property combinations. For example, we can see that a β ≥5 (lime green lines) can be achieved from both a moderately porous ($\phi$ =40%) targets with moderate internal friction (fi=1.35) and low cohesive strength (Yi0 = 1 kPa) or a target with no porosity ($\phi$ =0), high internal friction (fi=1.8), and moderate strength (Yi0= 100 kPa). In general, these plots showcase that material properties are not independent of one another in their effects on β. Additional trends can also be seen. Notably, simulations that had lower coefficients of internal friction (fi) and cohesion (Y0) resulted in higher values for β (represented by lighter and yellow lines); this trend held across a variety of porosities ($\varphi$). Similarly, higher cohesion and coefficients of friction resulted in lower beta (blue and purple). This trend was consistent across the codes used by the IWG. Figure 12 illustrates that material porosity ($\varphi$), coefficient of internal friction (fi), and cohesion (Yi0 or Y0) are all important in the cratering process and the momentum enhancement factor calculations, but the relationship is complex.

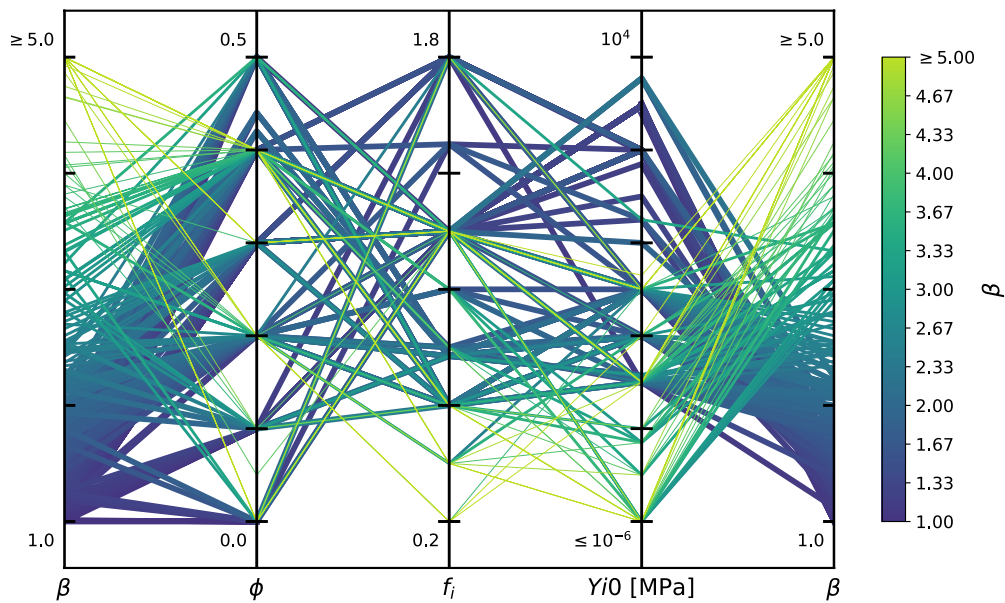

Figure 12. Parallel-line plot showing the interaction of material strength parameters (fi, Yi0) and porosity ($\phi$) on the momentum enhancement factor (β). Each line represents a single simulation and is colored by the calculated β.

The thickness of the line is used as a visualization tool to allow multiple overlapping lines to be shown. The vertical axes show the material property inputs for each simulation: porosity ($\varphi$), coefficient of friction (fi) and material strength/cohesion (Yi0 (also called Y0 in some codes)). These three material parameters significantly affect momentum enhancement, but their relationship is complex, though general trends can be seen.

Even if multiple combinations of material properties resulted in similar momentum enhancements, these impacts produced different crater morphologies (Fig. 13). For example, simulations suggested an impact into a strong (1 MPa), homogeneous, non-porous surface resulted in the same momentum enhancement as an impact into a weak (1 kPa), 50% porous surface ($\beta \sim 2.7$). However, these two impacts produced craters that differed considerably in size, with diameters of ~6.5 m and ~20.5 m, respectively. Another factor that can introduce similar beta values from different initial conditions is target layering (Fig. 13). Indeed, inferring layering of the target from the Deep Impact experiment led to low ejection angles (which would translate to a lower $\beta$) for the DI vapor plume suggested that the impact was well coupled to the target surface and indicated a denser (more competent) layer closer to the surface and a nested crater (Schultz et al. 2007, Schultz et al. 2013). Similar to this, for the DART case, impacts that result in similar values of $\beta$ could result in different crater morphologies, depending on the pre-impact upper-layer thickness (Quaide and Oberbeck 1968) or porosity. When Hera reaches the Didymos system, additional constraints could be placed on Dimorphos's material properties based on the data Hera collects.

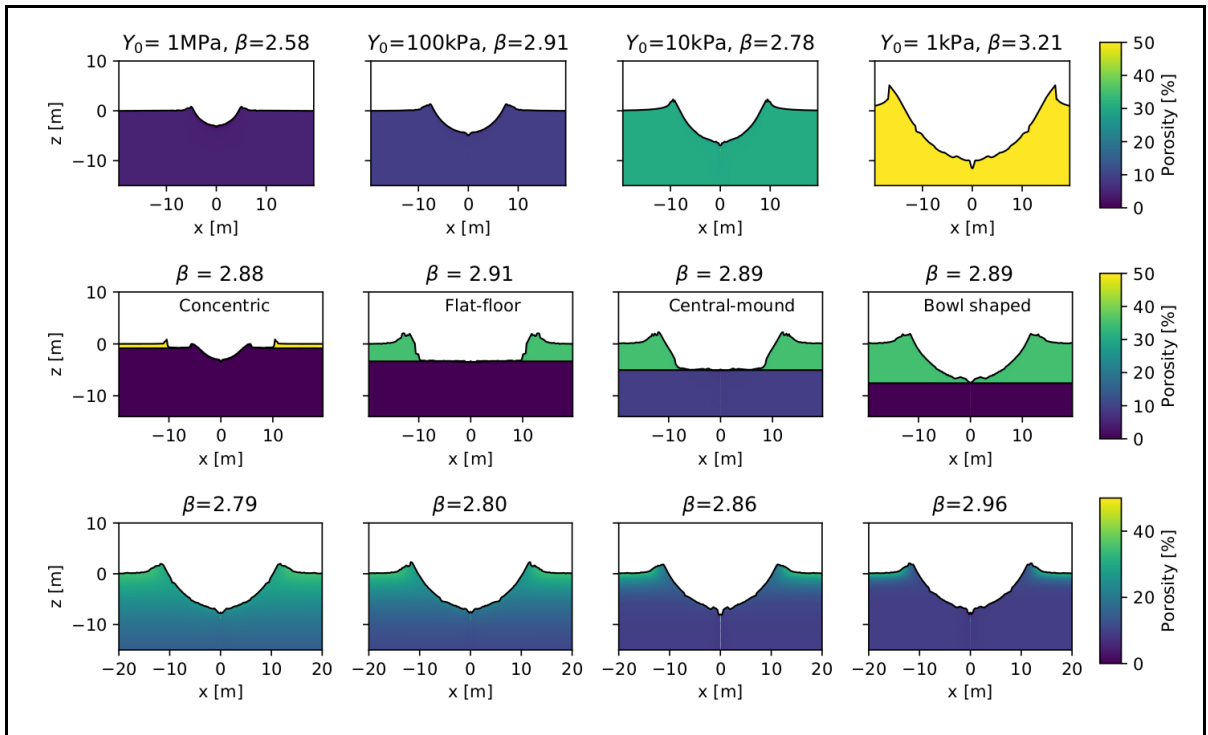

Fig. 13: Crater morphologies from iSALE-2D impact simulations of DART-like impacts into targets with a variety of material properties and structures that produce very similar $\beta$ values.

Because little is known about Dimorphos at this stage of the mission, determining a value for β is non-trivial. Thus, the impact simulations will be vital to determining the momentum enhancement. The knowledge gained from the simulations described in this work will provide the foundation for determining what the material properties of Dimorphos are likely to be based on observations. These hypotheses will be augmented from images from LICIAcube to help place constraints on Dimorphos's material properties (see Section 4.1) to provide estimates for deflection efficiency.

## 4.3    What Can We Expect from the DART Impact?

The discussions in this paper provided information about what could be expected following the DART impact for a variety of target (Dimorphos) conditions. These conditions include impact geometry parameters related to the asteroid slope and local tilts as well as target property parameters that may affect the cratering process and thus the expected deflection velocity and momentum enhancement. In order to best predict what the results of the DART impact may be, then, understanding the potential target properties is essential.

Some of the properties of interest, and their current best estimates (DART project DRA, January 2022), appear in Table 3.

Table 3. Impact parameters and properties of Dimorphos that may affect the DART impact

| Property of Interest | DART project DRA value | Notes |
|---|---|---|
| Impact velocity | 6.14 km s$^{-1}$ | |
| Slope of Boulder size frequency distribution | -3.5 | Assumes same sfd as Itokawa (Michikami et al. 2008) |
| Density (kg m$^{-3}$) | 2170 (+/- 350) | Leads to ~20% porosity (Richardson et al, 2022) |
| Possible Cohesion | 10 Pa - ? | See Section 3.1, based on dynamical considerations for Didymos |

As described in previous sections, numerous studies provided information about what may be expected following the DART impact into Dimorphos. Two of the most important variables affecting deflection are plotted in Figure 14 (porosity and cohesion). The points represent values determined from individual simulations; occasionally, multiple simulations overlapped in parameter space, and in that case, the larger predicted β value is represented by the colored circle. With this information, we can use two end-member possibilities for material parameters based on our understanding of Dimorphos to better understand what values of β may be expected. Dynamical considerations suggest that the strength of Dimorphos could be around 10 Pa, while density and volume estimates provide a porosity estimate of ~20%. This combination of material properties could result in a β value of ~5 (Figure 14). Alternatively, if Dimorphos possesses higher material strength (we assume 1 and 10 MPa, roughly based on Housen and Holsapple 2011), β~2. If Dimorphos turns out to be a remnant boulder with strength of intact rock (~> 100 MPa (e.g., Cotto-Figueroa 2016)), β could be as low as 1.

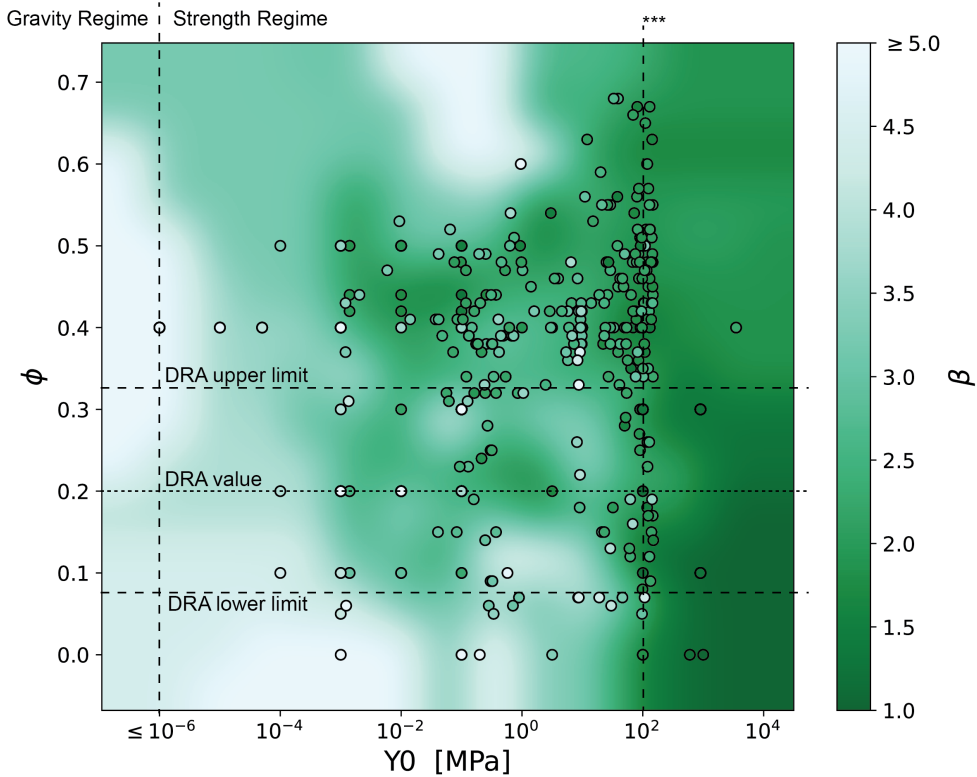

Figure 14: A summary of simulation results from the IWG showing how β is a function of assumed yield strength (cohesion, Y0) and target porosity (φ). Individual simulations are plotted as circles and colored by the calculated value for β. In cases where more than one simulation is plotted in the same parameter space, the simulation with the highest β is plotted on top. Dashes horizontal lines are plotted for the range of porosity values possible based on the current DART project Design Reference Asteroid (DRA), the "DRA" value represents the currently assumed most likely porosity. The vertical lines show representative potential strength values, including the transition where crater scaling changes from gravity to strength. Dynamical considerations suggest Dimorphos could be as weak as 10 Pa. The line at *** represents the strong end member for meteorite strengths calculated by Cotto-Figueroa (2016).

## 5 CONCLUSIONS AND PREDICTIONS FOR DART

The Hera mission will provide detailed measurements of both the DART impact outcome in terms of the crater's properties and momentum transfer as well as the properties of Dimorphos that have the greatest influence on the response of the object to an impact (Michel et al. 2022). In particular, the low-frequency radar JuRA onboard the Juventas Cubesat will provide the first measurements of the subsurface and internal properties of an asteroid. This can be added to information about the impact conditions that will be provided by DART and the first instant after the impact that will be imaged by LICIACube. Measurements of the physical and compositional properties of the target together with the high-resolution images and measurements of the DART impact crater by Hera will provide sufficient information that will allow for a robust validation of impact simulations at the scale of an asteroid. Such validation is crucial to greatly increase confidence in the numerical predictions of kinetic impact deflection and to extrapolate the knowledge gained by the AIDA cooperation to other scenarios.

The AIDA/DART IWG evaluated the effects of a variety of impact conditions and target properties on crater formation, ejecta properties, and momentum enhancement (including deflection velocity) in preparation for the DART impact. These studies provided important information on what may be expected from a kinetic impactor for different potential asteroid conditions. Following the DART impact, the knowledge gained from these studies will be used to inform the initial post-impact simulations recreating the conditions during the impact, including providing estimates for potential material properties of Dimorphos and the momentum enhancement factor resulting from DART's impact.

Some key takeaway points from these studies and important considerations are as follows:
- Numerical simulations are a vital component of understanding the momentum enhancement following the DART impact. The period change will be measured by ground-based observations, and deflection velocity (in the orbital velocity direction) can then be estimated. Impact simulations will be used to assess the material properties and impact parameters that are most likely to cause the observed changes. From these observations, an estimate for $\beta$ will be provided.
- Deflection velocity is directly calculated by numerical simulations. This parameter is most easily compared to observed period change. A benefit of impact simulations is the ability to examine the 3-dimensional nature of the impact process, and any out-of-plane effects resulting from the impact can be identified, including out-of-plane deflection velocity and ejecta. Impact simulations will be used to identify specific offsets ($\epsilon$) in the definition of $\beta$ from Rivkin et al. (2022).
- Material properties affect crater size, ejecta processes, and the resulting deflection velocity and momentum enhancement following the

DART impact. Numerical studies suggest that porosity and cohesion of the material are the two material properties with the largest effect on $\beta$ and deflection velocity.

- Ejecta mass and velocity profiles can be predicted by numerical simulations and will be estimated following the DART impact based on observed effects. Ejection angle and ejecta velocity are also sensitive to material properties, thus LICIAcube images, compared with impact simulations, will be used to place constraints on potential strength and porosity values for Dimorphos.

- Simulations assuming DART is a spherical impactor (with equivalent mass to the expected spacecraft mass at impact) overestimate the momentum enhancement $\beta$ by 5-20% and the crater size by 79-147% compared to simulations of the entire complex spacecraft. A simplified impactor geometry of three-spheres provides reasonable estimates for $\beta$ and crater size while preserving computational efficiency.

- Multiple combinations of material properties (e.g., strength parameters and/or porosity/subsurface structure) can result in similar $\beta$ values. Impact simulations coupled with direct observations can be used to identify reasonable estimates for Dimorphos's material properties that provide deflection velocities matching the observed period change. These data, coupled with mass and volume estimates from the shape model generated by DART observations, will provide an estimate for the momentum enhancement resulting from the DART impact.

- Extrapolation of results from impact experiments suggest a $\beta$ of at least 3, and perhaps much larger, for DART. However, it is important to note that these experiments did not include the full complexity of the DART spacecraft, and thus may also overestimate $\beta$.

- The local geology at the point of impact (e.g., presence of boulders, local tilt) will affect the crater formation and deflection processes. Images from DART will provide important information about the conditions at, and near, the impact site, which will allow follow-on simulations to adequately account for these uncertainties.

- The DART impact will not catastrophically disrupt Dimorphos. In the end-member of a fully strengthless-target (which is unrealistic), DART is still in the sub-catastrophic regime. However, if the DART impact does occur in the gravity regime, there is the potential for such a larger crater to form that it could lead to reshaping Dimorphos. In this regime, most of the crater growth could be gravity controlled until very late stages, when weak resistance (cohesion, internal friction) arrests excavation.

- While unknown, estimates for reasonable potential material properties of Dimorphos provide predictions for momentum transfer efficiency of 1-5, depending on end-member cases in the strength regime.

## 6 ACKNOWLEDGMENTS

This work was supported by the DART mission, NASA Contract No. 80MSFC20D0004. This work has received funding from the European Union's Horizon 2020 research and innovation program under grant agreement No. 870377 (project , NEO-MAPP) and from the French space agency CNES and the CNRS through the MITI interdisciplinary programs. We gratefully acknowledge the developers of iSALE-2D (www.isale-code.de), including Dirk Elbeshausen, Boris Ivanov and Jay Melosh. This work was supported by the Italian Space Agency (ASI) within the LICIACube project (ASI-INAF agreement AC n. 2019-31-HH.0; C.B. and C.M.S. appreciate support by the German Research Foundation (DFG) project 39848852. Alice Lucchetti, Maurizio Pajola (and all other coauthors that will be from LICIAcube) team acknowledges financial support from Agenzia Spaziale Italiana (ASI, contract No. 2019-31-HH.0). JO and IH were supported by the Spanish State Research Agency (AEI) Project No. MDM-2017-0737 Unidad de Excelencia "María de Maeztu" — Centro de Astrobiología (CSIC-INTA). They are also grateful for all logistical support provided by Instituto Nacional de Técnica Aeroespacial (INTA). Parts of this work were performed under the auspices of the U.S. Department of Energy by Lawrence Livermore National Laboratory under Contract DE-AC52- 07NA27344. LLNL-JRNL-831475. This work was supported in part by a Chick Keller Postdoctoral Fellowship through the Center for Space and Earth Science at Los Alamos National Laboratory. Parts of this work were supported by the Advanced Simulation and Computing (ASC)—Threat Reduction and ASC—Planetary Defense programs. Los Alamos National Laboratory, an affirmative action/equal opportunity employer, is operated by Triad National Security, LLC, for the National Nuclear Security Administration of the U.S. Department of Energy under contract 89233218NCA000001. LA-UR-22-21164

Appendix A. Hydrocode Validation Studies Relevant to Planetary Defense

Numerical simulations must be validated against experimental and theoretical data to ensure they predict accurate outcomes. For DART, specific comparisons against relevant planetary defense problems are useful to ensure that the specific codes being used by the IWG are robust for planetary defense problems. Here, we briefly summarize some validation studies the IWG undertook for the various codes used by the team.

Validation Studies

The DART impact occurs in the hypervelocity cratering regime, which is a complex process that requires shock physics codes for numerical modeling. The amount by which the asteroid Dimorphos can be deflected is highly dependent on its target properties and structure. Such an analysis based on numerical models requires accurate validation of the applied numerical codes. Previously, our numerical codes and the underlying material models (e.g., pore compaction, target behavior under high pressures) have been validated against a range of laboratory-based experiments.

For example, for the Bern SPH code, the validation tests include impact experiments with non-porous (Benz and Asphaug 1995) and porous targets (Jutzi et al. 2009, Jutzi 2015) and granular flow experiments (Jutzi, 2015). Recent validation simulations reproduced the ejecta velocity distributions resulting from impacts into frozen clay targets with varying strength and friction properties (Arakawa et al. 2022).

The iSALE shock physics code was validated against a range of experiments in terms of the crater size in metal targets (e.g., Davison et al. (2011) for impacts into aluminum), competent rock targets (e.g., by Güldemeister et al. (2015) and Winkler et al (2018) for non-porous quartzite, marble, and porous sandstone) and granular targets (e.g., Ormö et al. (2015) and Wünnemann et al. (2016) for impacts into quartz sand). Wünnemann et al. (2016) also validated the simulation results against the final ejecta deposit. Validations focusing on material ejection have been conducted by Luther et al. (2018) and Raducan et al. (2019) for different granular materials.

However, most previous experiments considered impact conditions and target materials different from what we expect to find on an asteroid's surface. In order to validate our models against appropriate impact conditions and materials, laboratory-based data from impact experiments that were specifically designed to mimic the assumed surface materials and structures of Dimorphos were needed. In this section, we summarize several validation and benchmarking campaigns undertaken in the context of the DART and Hera missions.

Chourey et al. (2020) conducted laboratory-scale impact experiments into lunar regolith simulant at velocities on the order of 2 km/s. The lunar regolith simulant is considered to be a good analogue for the regolith material found on the surface of some asteroids (Sullivan et al. 2002). The experimental set-up allowed measurements of both the final crater size and the momentum deflection efficiency ($\beta$). We used these experiments to validate our numerical codes for impacts into homogeneous target materials. We reproduced these experiments with the grid-based iSALE-2D and the SPH codes Bern SPH and miluphcuda. Our simulation results are generally in good agreement with the experimental data and also among the different codes considered here (Luther et al., 2021). The miluphcuda SPH code has previously been validated against laboratory-scale high velocity impacts into solid and brittle materials as well as granular flow experiments (Schäfer et al. 2016, Schäfer et al. 2020).

Ormö et al. (2022) carried out impact experiments into targets specifically designed to reproduce rubble-pile asteroids' surface materials and structures. The experiments were performed at the Experimental Projectile Impact Chamber (EPIC) at Centro de Astrobiología CSIC-INTA, Spain, and used a quarter-space set-up (Ormö et al. 2015). Ormö et al. (2020) launched 20-mm delrin projectiles at velocities ~400 m/s into sand targets with different configurations of embedded, porous, ceramic "boulders" of similar size and mass as the projectiles. The ceramic material of these "boulders" is considered to be a good mechanical analog for boulders found on the rubble-pile asteroids Ryugu and Bennu (Ballouz et al. 2020). The experiments were closely reproduced using the Bern SPH code and gave information on the effects of embedded boulders on crater size, shape, and material displacement as well as ejection mechanisms for both boulders and the sand matrix, thus allowing the validation and calibration of the codes in order to face full-size impact simulations such as DART.

Stickle et al. (2020) included validation results for CTH and Spheral against a hypervelocity impact experiment performed at the NASA Ames Vertical Gun Range. Both CTH and Spheral performed well in the blind-comparison to an AVGR experiment. Predictions of crater size and extent of fracturing were between 10 and 40% different than the measured values. Comparisons of Spheral against impact experiments on a basalt target (Nakamura and Fujiwara 1991) also show that the simulations are sensitive to the selected strain models, strength models, and material parameters. When appropriate choices for these models are used in conjunction with well-constrained material parameters, the simulations closely resemble the experimental results (Remington et al. 2020).

The FLAG hydrocode has been verified against a 1D analytic solutions in 1D, 2D, and 3D simulations of impacts in both the strength-dominated and gravity-dominated regimes (Pierazzo et al. 1997; Caldwell et al. 2018). FLAG was also validated against laboratory impact experimental data (Caldwell et al. 2018). The study indicated the mesh resolution can result in over- or underestimations of maximum pressure ranging from -4.15% to 4.15% in strength-dominated 2D simulations. FLAG appeared to converge at a resolution of about 10 cells per projectile radius (10 cppr), reducing the computational resources of a fully resolved 40 cppr simulation from about 28 hours to about 25 minutes with little variation in results (Caldwell et al. 2018). In 3D, the coarser resolution required for the large computational domain resulted in overestimations of about 10.79% in the strength-dominated regime (Caldwell et al. 2018). In 2D axisymmetric simulations of a laboratory impact experiment, FLAG overestimated crater depth by about 2.44% and underestimated crater radius by about 6.2%. The study indicated the depth overestimation was likely attributed to the combination of an axisymmetric boundary along the impact trajectory as well as the existence of gravity in the simulation (Caldwell et al. 2018). Other impact studies in FLAG indicated EOS variations had little effect (Caldwell et al. 2018, Caldwell 2019). Across 5 tested constitutive models, the maximum pressure varied by, at most, 0.12 GPa, corresponding to deviations of about 0.21% (Caldwell 2019).